\documentclass[aoas,preprint]{imsart}

\RequirePackage[OT1]{fontenc}
\RequirePackage{amsthm,amsmath}
\RequirePackage{natbib}
\RequirePackage[colorlinks,citecolor=blue,urlcolor=blue]{hyperref}

\usepackage{graphicx,psfrag,epsf}
\usepackage{amsmath,amsfonts,amsthm}
\usepackage{enumerate}
\usepackage{color}
\usepackage{amssymb}
\usepackage{array}
\usepackage{dsfont}
\usepackage{xcolor}
\usepackage{booktabs}
\usepackage{tabularx}
\usepackage[font = small,labelfont=bf,textfont=it]{subcaption}
\usepackage[font = small,labelfont=bf,textfont=it]{caption} 
\usepackage{textcomp}
\usepackage{algorithm}
\usepackage[noend]{algpseudocode}
   
\newtheorem{theorem}{Theorem}
\newtheorem{proposition}[theorem]{Proposition}
\newtheorem{corollary}{Corollary}

\makeatletter

\newcommand{\distas}[1]{\mathbin{\overset{#1}{\kern\z@\sim}}}
\newcommand{\bm}[1]{\mathbf{#1}}
\newsavebox{\mybox}\newsavebox{\mysim}
\newcommand{\distras}[1]{%
  \savebox{\mybox}{\hbox{\kern3pt$\scriptstyle#1$\kern3pt}}%
  \savebox{\mysim}{\hbox{$\sim$}}%
  \mathbin{\overset{#1}{\kern\z@\resizebox{\wd\mybox}{\ht\mysim}{$\sim$}}}%
}
\DeclareMathOperator*{\argmin}{\arg\!\min}

\renewcommand{\fnum@algorithm}{\fname@algorithm}

\makeatother

\startlocaldefs
\numberwithin{equation}{section}
\theoremstyle{plain}

\endlocaldefs

\begin{document}

\begin{frontmatter}
\title{Adaptive design for Gaussian process regression under censoring\thanksref{T1}}
\runtitle{Adaptive design under censoring}
 \runauthor{J. Chen et al.}

\thankstext{T1}{This work is supported by NSF CSSI Frameworks 2004571, NSF CMMI grant 1921646, and Piedmont Heart Institute.}

\begin{aug}

\author[A]{\fnms{Jialei} \snm{Chen}\ead[label=e1]{jialei.chen@gatech.edu}},
\author[B]{\fnms{Simon} \snm{Mak}\ead[label=e2]{sm769@duke.edu}},
\author[A]{\fnms{V. Roshan} \snm{Joseph}\ead[label=e3]{roshan@gatech.edu}}, 
\and
\author[A]{\fnms{Chuck} \snm{Zhang}\ead[label=e4]{chuck.zhang@gatech.edu}}

\address[A]{H. Milton Stewart School of
Industrial \& Systems Engineering
Georgia Institute of Technology,\\ \printead{e1,e3,e4}}

\address[B]{Department of Statistical Science, Duke University, \\
\printead{e2}}
\end{aug}

\begin{abstract}   

A key objective in engineering problems is to predict an unknown experimental surface over an input domain. In complex physical experiments, this may be hampered by response censoring, which results in a significant loss of information. 
For such problems, experimental design is paramount for maximizing predictive power using a small number of expensive experimental runs. To tackle this, we propose a novel adaptive design method, called the integrated \textit{censored} mean-squared error (ICMSE) method. The ICMSE method first estimates the posterior probability of a new observation being censored,
then adaptively chooses design points that minimize predictive uncertainty under censoring.
Adopting a Gaussian process regression model with product correlation function, the proposed ICMSE criterion is easy to evaluate, which allows for efficient design optimization. We demonstrate the effectiveness of the ICMSE design in two real-world applications on surgical planning and wafer manufacturing.

\end{abstract}

\begin{keyword}[class=MSC]
\kwd{62K20}
\end{keyword}

\begin{keyword}
\kwd{Adaptive sampling}
\kwd{Censored experiments}
\kwd{Experimental design}
\kwd{Kriging}
\kwd{Multi-fidelity modeling}
\end{keyword}

\end{frontmatter}

\section{Introduction}

\begin{figure}[!t]
\centering
\includegraphics[width=0.85\textwidth]{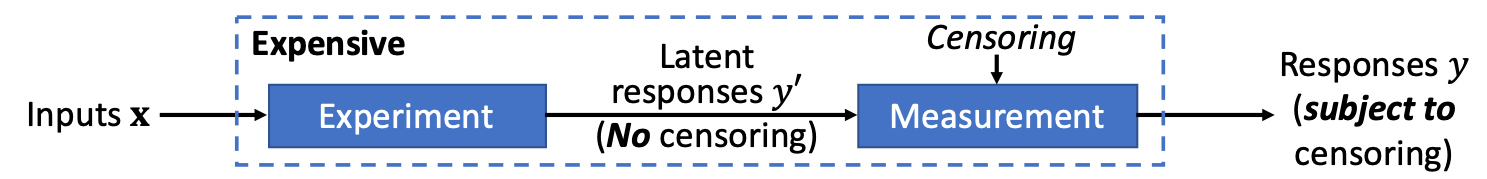}
\caption{\label{Fig:censor} An illustration of response censoring in the measurement process. The goal is to predict the response surface for the experiment prior to censoring.}
\end{figure}

In many engineering problems, a key objective is to predict an unknown experimental surface over an input domain.
However, for complex physical experiments, one can encounter  \textit{censoring}, i.e., the experimental response is missing or partially measured.  
Censoring arises from a variety of practical experimental constraints, including limits in measurement devices, safety considerations of experimenters, and a fixed experimental time budget. 
Fig \ref{Fig:censor} provides an illustration: 
in such cases, the experimental response of interest is latent, and the observed measurement is subject to censoring.
Here, censoring can result in significant loss of information, which leads to poor predictive performance \citep{brooks1982loss}. For example, suppose an engineer wishes to explore how pressure in a nuclear reactor changes under different control settings.
Due to safety concerns, experiments are forced to stop if the pressure hits a certain upper limit, leading to censored responses.
To further complicate matters, the input region which results in censoring is typically \textit{unknown} prior to experiments, and needs to be estimated from data.

Given the presence of censoring in physical experiments, it is therefore of interest to carefully design experimental runs, to best model and predict the experimental response surface.
We present a new integrated \textit{censored} mean-squared error (ICMSE) method, which sequentially selects physical experimental runs to minimize predictive uncertainty under \textit{censoring}. ICMSE leverages a Gaussian process model (GP; \citealp{sacks1989design}) -- a flexible Bayesian nonparametric model -- for the response surface, to obtain an easy-to-evaluate design criterion that maximizes GP's predictive power under censoring. We consider two flavors of ICMSE.
The first is a ``single-fidelity'' ICMSE method for sequentially designing (potentially) censored physical experiments. The second is a ``bi-fidelity'' ICMSE method for sequentially designing (potentially) censored physical experiments given auxiliary computer simulation data. 
The two settings are motivated from the following two applications.

\subsection{3D-printed aortic valves for surgical planning} \label{Sec:MotiExampleMM}
\begin{figure}[!t]
\centering
\includegraphics[width=0.99\textwidth]{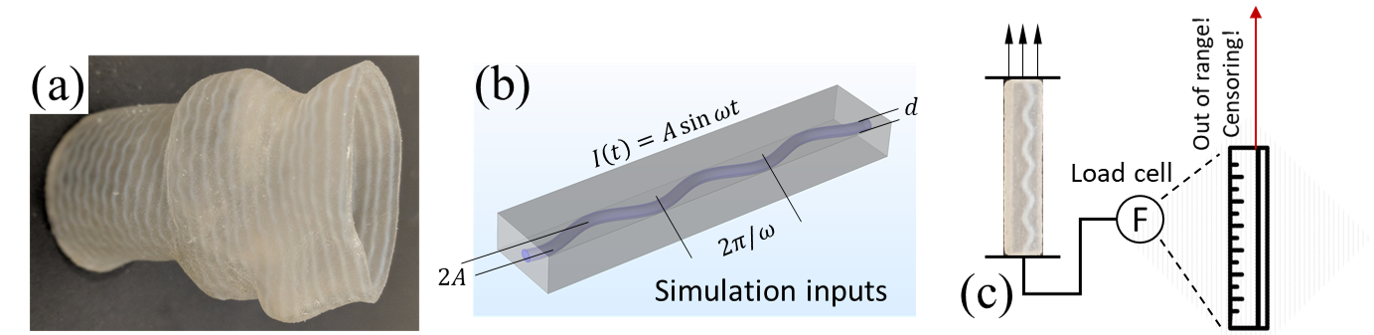}
\caption{\label{Fig:MMIntro} Illustrating the surgical planning application: (a) a 3D-printed aortic valve with enhanced  metamaterial, (b) simulation inputs in the computer experiment, (c) visualizing the physical experiment and the measurement censoring of the load cell (labeled ``F'').}
\end{figure}

The first motivating problem concerns the design of 3D-printed tissue-mimicking aortic valves for heart surgeries.
With advances in additive manufacturing \citep{gibson2014additive}, 3D-printed medical prototypes \citep{rengier20103d} 
play an increasingly important role in pre-surgical studies \citep{qian2017quantitative}.
They are particularly helpful in complicated heart diseases, e.g., aortic stenosis, where 3D-printed aortic valves can be used to select the best surgical option with minimal post-surgical complication \citep{chen2018generative}.
The printed aortic valve (see Fig \ref{Fig:MMIntro}(a)) contains a biomimetic substructure: an enhancement polymer (white) is embedded in a substrate polymer (clear); this is known as \textit{metamaterial} \citep{wang2016dual} in the materials engineering literature.
The goal is to build a model to understand how the \textit{stiffness} of the metamaterials is affected by the \textit{geometry} of the enhancement polymer (see Fig \ref{Fig:MMIntro}(b)).
This model can then be used by doctors to select a polymer geometry that mimics the target stiffness of the specific patient -- a procedure known as ``tissue-mimicking'' \citep{chen2018JASA}. An accurate tissue-mimicking is paramount for surgery success, since inaccurate stiffness may lead to severe post-surgery complications and death.

Using earlier terminology, this is a bi-fidelity modeling problem involving two types of experiments: a pre-conducted database of computer simulations and patient-specific physical experiments.
The physical experiments here are very \textit{costly}: we need to 3D print each metamaterial sample, then physically test its stiffness using a load cell. Furthermore, the measurement from physical experiments may be \textit{censored} due to an inherent upper limit of the testing machine. This is shown in Fig \ref{Fig:MMIntro}(c): if the metamaterial sample is stiffer than the load cell (i.e., a spring), the experiment is forced to stop to prevent breakage of the load cell.
One workaround is to use a stiffer load cell, however, it is oftentimes \textit{not} a preferable option: a stiffer load cell with a broader measurement range can be very expensive, costing over a hundred times more than the standard integrated load cells.
Here, the proposed ICMSE method can adaptively design experimental runs to maximize the predictive power of a GP model under censoring. We show later in Section \ref{Sec:MMResult}, ICMSE can lead to better predictions for younger patients (with stiff tissues, which can be \textit{censored} in experiments) and older patients (with soft tissues, which are \textit{uncensored} in experiments) \citep{sicard2018aging}.
This then leads to greatly improved tissue-mimicking performance for personalized printed valves \citep{chen2018JASA}, which is crucial for improving heart surgery success rate \citep{qian2017quantitative}.
Our method is particularly valuable in urgent heart surgeries, where one can perform only a small number of runs prior to the actual surgery.

\subsection{Thermal processing in wafer manufacturing} \label{Sec:MotiExampleLH}
\begin{figure}[!t]
\centering
\includegraphics[width=0.7\textwidth]{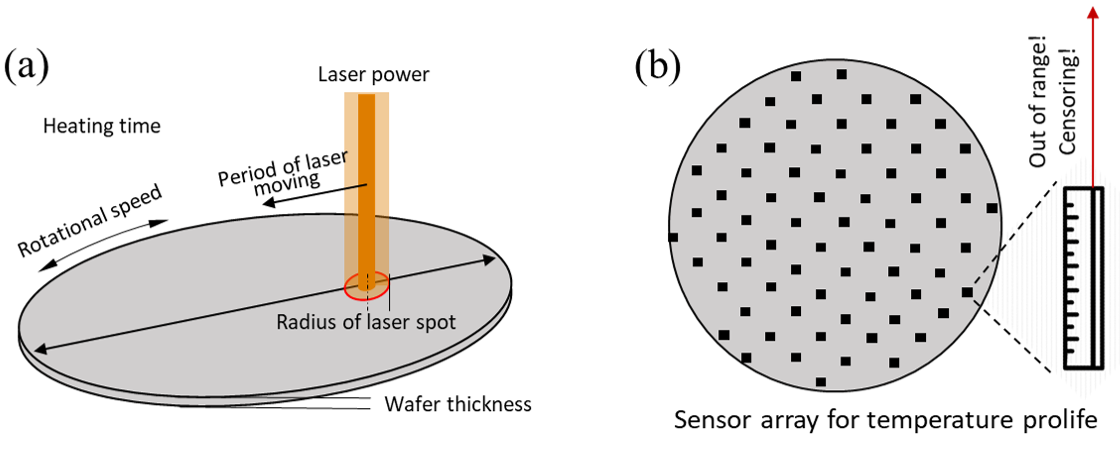}
\caption{\label{Fig:LHIntro} Illustrating the wafer manufacturing application: (a) visualizing the thermal processing procedure with the 6 input parameters, (b) visualizing the measurement censoring of the temperature sensor array.}
\end{figure}

The second problem considers the design of the semiconductor wafer manufacturing process \citep{quirk2001semiconductor,jin2012sequential}. Wafer manufacturing involves processing silicon wafers in a series of refinement stages, to be used as circuit chips.
Among these stages, thermal processing is one of the most important stages \citep{singh2000rapid},
since it facilitates the necessary chemical reactions and allows for surface oxidation. 
Fig \ref{Fig:LHIntro}(a) illustrates the typical thermal processing procedure: a laser beam (in orange) is moved back and forth over a rotating wafer. The output of interest here is the \textit{minimal} temperature over the whole wafer after heating; a higher minimal temperature facilitates better completeness of chemical reactions, which leads to better quality of the final wafer product \citep{goodson1993annealing,van1998time}. However, higher temperatures may result in higher energy costs for heating. With the fitted predictive model on minimal wafer temperature, industrial engineers can then use this to optimize a heating process which is economical (i.e., conserves heating power) but also meets target quality requirements.

However, laser heating experiments are quite costly, involving high material and operation costs. In industrial settings, the minimal wafer temperature is often subject to \textit{censoring}, due to the nature of measurement procedures.
This is shown in Fig \ref{Fig:LHIntro}(b): the wafer temperature is typically measured by either an array of temperature sensors or a thermal camera, both of which have upper measurement limits \citep{feteira2009negative}. The minimal temperature is censored when the whole sensor array reaches the measurement limits. While more sophisticated sensors exist, they are much more expensive and may lead to tedious do-overs of experiments.
The proposed single-fidelity ICMSE method can be used to adaptively design experimental runs that maximize the predictive power of a GP model under censoring. We show later in Section \ref{Sec:LHresult} that the resulting model using ICMSE enjoys improved predictive performance for high wafer temperatures (that are potentially censored) and low temperatures (that are not censored), to ensure flexibility for different quality requirements. The fitted model can then be used to find an optimal thermal processing setting, which minimizes operation costs subject to target quality requirements.

\subsection{Literature} 
GP regression (or \textit{kriging}, see \citealp{matheron1963principles}) is widely used as a predictive model for expensive experiments \citep{sacks1989design}, and has been applied in cosmology \citep{kaufman2011efficient}, aerospace engineering \citep{mak2018efficient}, healthcare \citep{chen2018JASA}, and other applications.
The key appeals of GPs are the flexible nonparametric model structure and closed-form expressions for prediction and uncertainty quantification \citep{santner2018design}. 
In the engineering literature, GPs have been used for modeling expensive physical experiments \citep{ankenman2010stochastic}, integrating computer and physical experiments \citep{KO2001}, and incorporating various constraints \citep{henkenjohann2005adaptive,groot2012gaussian,da2012gaussian,lopez2018finite,ding2019bdrygp}.
We will adapt in this work a recent censored GP model \citep{cao2018model}, which integrates censored physical experimental data.

There have been several works in the literature on experimental design under response censoring, see, e.g., \cite{borth1996optimal,monroe2008experimental}. 
These methods, however, presume a parametric form for the response surface, which may be a dangerous assumption for black-box experiments, hence the recent shift for more nonparametric models such as GPs.
Existing design methods for GPs can be divided into two categories -- space-filling and model-based designs. 
Space-filling designs aim to fill empty gaps in the input space; this includes minimax designs \citep{johnson1990minimax}, maximin designs  \citep{morris1995exploratory}, and maximum projection designs \citep{joseph2015maximum}. Model-based designs instead maximize an optimality criterion based on an \textit{assumed} GP model; this
includes integrated mean-squared error designs \citep{sacks1989design} and maximum entropy designs \citep{shewry1987maximum}. Such designs can also be implemented sequentially in an adaptive manner, see  \cite{lam2008sequential,xiong2013sequential,chen2017sequential,bect2019supermartingale}.
Recently, \cite{binois2019replication} proposed a design method for a heteroscedastic GP model (i.e., under input-dependent noise); this provides a flexible framework that allows for different correlation functions, closed-form gradients for optimization, and batch sequential implementation.

The above GP design methods, however, do not consider potential response \textit{censoring}.
The key challenge in incorporating censoring information is that an experimenter does not know which inputs may lead to censoring prior to experimentation, since the response surface is black-box.
The proposed ICMSE method addresses this by leveraging a GP model on the unknown response surface: it first estimates the posterior probability of a potential observation being censored, and then finds design points that minimize predictive uncertainty under censoring. Under product correlation functions, our method admits an easy-to-evaluate design criterion, which allows for efficient sequential sampling. We show that ICMSE can yield considerably improved predictive performance over existing design methods (which do not consider censoring), in both motivating applications.

\subsection{Structure}
Section \ref{Sec:SingeF} presents the ICMSE design method for the single-fidelity setting, with only physical experiment data.
Section \ref{Sec:MultiF} extends the ICMSE method for the bi-fidelity setting, where auxiliary computer simulation data are available.
Section \ref{Sec:CS} demonstrates the effectiveness of ICMSE in the two motivating applications. Section \ref{sec:conclusion} concludes the work.

\section{ICMSE design} \label{Sec:SingeF}

We now present the ICMSE design method for the single-fidelity setting; a more elaborate bi-fidelity setting is discussed later in Section \ref{Sec:MultiF}. We first review the GP model for censored data, and derive the proposed ICMSE design criterion. We then visualize this via a 1-dimensional (1D) example, and provide some insights.

\subsection{Modeling framework} \label{Sec:PEModel}
We adopt the following model for physical experiments. Let $\bm{x}_i \in [0,1]^p$ be a vector of $p$ input variables (each normalized to $[0,1]$), and let $y_i'$ be its latent response from the physical experiment \textit{prior to} potential censoring (see Fig \ref{Fig:censor}).
We assume:
\begin{align}
y_i'=\xi(\bm{x}_i)+\epsilon_i, \quad i = 1,2, \cdots, n,
\label{Equ:PEModel}
\end{align}
where $\xi(\bm{x}_i)$ is the mean of the latent response $y'_i$ at input $\bm{x}_i$, and $\epsilon_i$ is the corresponding measurement error. Since $\xi(\cdot)$ is unknown, we further assign to it a GP prior with mean $\mu_\xi$, variance $\sigma^2_\xi$, and correlation function $R_{\boldsymbol{\theta}_\xi}(\cdot,\cdot)$ with parameters $\boldsymbol{\theta}_\xi$. This is denoted as:
\begin{equation}
\xi(\cdot) \sim \text{GP} \left(\mu_{\xi}, \sigma^2_\xi R_{\boldsymbol{\theta}_\xi}(\cdot, \cdot) \right).
\label{Equ:PEGP}
\end{equation}
The experimental noise $\epsilon_i \distas{i.i.d.} \mathcal{N}(0,\sigma^2_\epsilon)$ is assumed to be i.i.d. normally distributed, and independent of $\xi(\cdot)$.

For simplicity, we consider only the case of right-censoring below, i.e., censoring of the response only when it exceeds some \textit{known} upper limit (this is the setting for both motivating applications). All equations and insights derived in the paper hold analogously for the general case of interval censoring, albeit with more cumbersome notation. Suppose, from $n$ experiments, $n_o$ responses are observed without censoring, and $n_c$ responses are right-censored at limit $c$, where $n_o + n_c = n$. 
The training set experimental data can then be written as the set $\mathcal{Y}_n=\{\bm{y}_{o}, \bm{y}_c' \geq \bm{c}\}$, where $\bm{y}_o$ is a vector of \textit{observed} responses at inputs $\bm{x}_{o} =\bm{x}_{1:n_o} =\{\bm{x}_1, \cdots, \bm{x}_{n_o}\}$, $\bm{y}_c'$ is the latent response vector for inputs in censored regions $\bm{x}_{c} =\bm{x}_{(n_o+1):n}$ prior to censoring, and $\bm{c} = [c,\cdots, c]^T$ is the vector of the right-censoring limit. 
Assuming known model parameters, a straightforward adaptation of the equations (11) and (12) in \cite{cao2018model} gives the following expressions for the conditional mean and variance of $\xi(\bm{x}_{\rm new})$ at new input $\bm{x}_{\rm new}$: 
\begin{align}
\hat{\xi}(\bm{x}_{\rm new}) = \mathbb{E}[\xi(\bm{x}_{\rm new})|\mathcal{Y}_n]&=\mu_\xi + \boldsymbol{\gamma}_{n,\rm new}^T \bm{\Gamma}_n^{-1}\left([\bm{y}_{o}, \hat{\bm{y}}_{c}]^T - \mu_\xi \cdot \textbf{1}_{n}\right),
\label{Equ:Censor_E}\\
s^2(\bm{x}_{\rm new}) =\text{Var} [\xi(\bm{x}_{\rm new})|\mathcal{Y}_n]&=\sigma^2_\xi-\boldsymbol{\gamma}_{n,\rm new}^T (\bm{\Gamma}^{-1}_n-\bm{\Gamma}_n^{-1}{\bm{\Sigma}}\bm{\Gamma}_n^{-1})\boldsymbol{\gamma}_{n,\rm new}.
\label{Equ:Censor_Var}
\end{align}
Here, $\bm{\Gamma}_n = \sigma^2_\xi{[R_{\boldsymbol{\theta}_\xi}(\bm{x}_i, \bm{x}_j)]_{i=1}^n} _{j=1}^n + \sigma^2_\epsilon \bm{I}_{n}$, $\boldsymbol{\gamma}_{n,\rm new}=\sigma^2_\xi\big[R_{\boldsymbol{\theta}_\xi}(\bm{x}_1, \bm{x}_{\rm new}), \cdots, \break R_{\boldsymbol{\theta}_\xi}(\bm{x}_n, \bm{x}_{\rm new}) \big]^T$,  $\boldsymbol{1}_n$ is a one-vector of length $n$, and $\bm{I}_{n}$ is an $n \times n$ identity matrix. Furthermore, $\hat{\bm{y}}_{c}=\mathbb{E}[\bm{y}_{c}' |\mathcal{Y}_n]$ is the expected response for the latent vector $\bm{y}_c'$ given the dataset $\mathcal{Y}_n$,
${\bm{\Sigma}}_{c}=\text{Var}[\bm{y}_c'|\mathcal{Y}_n]$ is its conditional variance, and ${\bm{\Sigma}} = \text{diag}(\textbf{0}_{n_{o}},{\bm{\Sigma}}_{c})$. The computation of these quantities will be discussed later in Section \ref{Sec:AdaAlgor}. The conditional mean \eqref{Equ:Censor_E} is used to predict the mean experimental response at an untested input $\bm{x}_{\rm new}$, and the conditional variance \eqref{Equ:Censor_Var} is used to quantify predictive uncertainty.

In the case of no censoring (i.e., $\mathcal{Y}_n = \{\bm{y}_o\}$), equations \eqref{Equ:Censor_E} and \eqref{Equ:Censor_Var} reduce to:
\begin{align}
\hat{\xi}(\bm{x}_{\rm new}) = \mathbb{E}[\xi(\bm{x}_{\rm new})|\mathcal{Y}_n]&=\mu_\xi + \boldsymbol{\gamma}_{n,\rm new}^T \bm{\Gamma}_n^{-1}\left(\bm{y}_o - \mu_\xi \cdot \textbf{1}_{n}\right), \quad \text{and}
\label{Equ:Uncensor_E}\\
s^2(\bm{x}_{\rm new}) =\text{Var} [\xi(\bm{x}_{\rm new})|\mathcal{Y}_n]&=\sigma^2_\xi-\boldsymbol{\gamma}_{n,\rm new}^T \bm{\Gamma}^{-1}_n \boldsymbol{\gamma}_{n,\rm new}.
\label{Equ:Uncensor_Var}
\end{align}
These are precisely the conditional mean and variance expressions for the standard GP regression model \citep{santner2018design}, which is as expected.

\subsection{Design criterion} \label{Sec:ICMSEFormula}
Now, given data $\mathcal{Y}_n$ from $n$ experiments ($n_o$ of which are observed exactly, $n_c$ of which are censored), we propose a new design method that accounts for the posterior probability of a potential observation being censored.
Let $\bm{x}_{n+1}$ be a potential next input for experimentation, $Y_{n+1}'$ be its latent response \textit{prior to} censoring, and $Y_{n+1} = Y_{n+1}'(1-\mathds{1}_{\{Y_{n+1}'\geq c\}}) + c \mathds{1}_{\{Y_{n+1}'\geq c\}}$ be its corresponding observation \textit{after} censoring, with $\mathds{1}_{\{\cdot\}}$ denoting the indicator function. The proposed method chooses the next input $\bm{x}^*_{n+1}$ as:
\begin{align}
\begin{split}
\bm{x}^*_{n+1} =& \argmin_{\bm{x}_{n+1}} \text{ICMSE}(\bm{x}_{n+1})\\
:=& \argmin_{\bm{x}_{n+1}} \int_{[0,1]^p} \mathbb{E}_{Y_{n+1}|\mathcal{Y}_n}\left[\text{Var}(\xi(\bm{x}_{\rm new})|\mathcal{Y}_n,Y_{n+1}) \right] \; d\bm{x}_{\rm new}.
\label{Equ:SeqICMSE}
\end{split}
\end{align}
The design criterion $\text{ICMSE}(\bm{x}_{n+1})$ can be understood in two parts. First, the term $\text{Var}(\xi(\bm{x}_{\rm new})|\mathcal{Y}_n,Y_{n+1})$ quantifies the predictive variance (i.e., mean-squared error, MSE) of the mean response at an untested input $\bm{x}_{\rm new}$, given both the training data $\mathcal{Y}_n$ and the potential observation $Y_{n+1}$. This is a reasonable quantity to minimize for design, since we wish to find which new input $\bm{x}_{n+1}$ can minimize predictive uncertainty. Second, note that this MSE term cannot be used directly as a criterion, since it depends on the potential observation $Y_{n+1}$, which is yet to be observed. One way around this is to take the conditional expectation $\mathbb{E}_{Y_{n+1}|\mathcal{Y}_n}[\cdot]$ (more on this below). 
Finally, the integral over $[0,1]^p$ yields the average predictive uncertainty over the entire design space.

The proposed criterion in \eqref{Equ:SeqICMSE} can be viewed as an extension of the sequential integrated mean-squared error (IMSE) design \citep{lam2008sequential, santner2018design} for the censored response setting. Assuming no censoring (i.e., $\mathcal{Y}_n = \{\bm{y}_o\}$), the sequential IMSE design chooses the next input $\bm{x}_{n+1}^*$ by minimizing:
\begin{align}
\min_{\bm{x}_{n+1}} \text{IMSE}(\bm{x}_{n+1}):= \min_{\bm{x}_{n+1}} \int_{[0,1]^p} \text{Var}(\xi(\bm{x}_{\rm new})|\mathcal{Y}_n,Y_{n+1}') \; d\bm{x}_{\rm new}.
\label{Equ:Seq_IMSE}
\end{align}
Note that, in the \textit{uncensored} setting, the MSE term $\text{Var}(\xi(\bm{x}_{\rm new})|\mathcal{Y}_n,Y_{n+1}')$ in \eqref{Equ:Seq_IMSE} does \textit{not} depend on the potential observation $Y_{n+1}'$, which allows the criterion to be easily computed in practice. However, in the \textit{censored} setting at hand, not only does this MSE term \textit{depend} on $Y_{n+1}'$, but such an observation may not be directly observed due to censoring. The conditional expectation $\mathbb{E}_{Y_{n+1}|\mathcal{Y}_n}[\cdot]$ in \eqref{Equ:SeqICMSE} addresses this by accounting for the posterior probability of censoring in $Y_{n+1}'$.

One attractive feature of the ICMSE criterion \eqref{Equ:SeqICMSE} is that it will be \textit{adaptive} to the experimental responses from data. The criterion \eqref{Equ:SeqICMSE} inherently hinges on whether the potential observation $Y_{n+1}$ is censored (i.e., $Y'_{n+1} \geq c$) or not (i.e., $Y_{n+1}' < c$), but this censoring behavior needs to be estimated from experimental data. Viewed this way, the ICMSE criterion can be broken down into two steps: it (i) estimates the posterior probability of a new observation being censored from data, and then (ii) samples the next point that minimizes the \textit{average} predictive uncertainty under censoring.
We will show how our method adaptively incorporates the posterior probability of censoring $Y_{n+1}$ for sequential design, in contrast to the existing IMSE method \eqref{Equ:Seq_IMSE}.

\subsubsection{No censoring in training data} \label{Sec:NoPriorCensor}
To provide some intuition, consider a simplified scenario with no censoring in the \textit{training} set, i.e., $\mathcal{Y}_n = \{\bm{y}_o\}$ (censoring may still occur for the new $Y_{n+1}$). In this case, the following proposition gives an explicit expression for the ICMSE criterion.
\begin{proposition}
\label{prop:1}
Suppose there is no censoring in training data, i.e., $\mathcal{Y}_n = \{\bm{y}_o\}$. 
Then the ICMSE criterion \eqref{Equ:SeqICMSE} has the explicit expression:
\begin{align}
&\text{\rm ICMSE} (\bm{x}_{n+1}) =\int_{[0,1]^p}\sigma_{\rm new}^2-h_c(\bm{x}_{n+1})\rho_{\rm new}^2(\bm{x}_{n+1})\sigma_{\rm new}^2\; d\bm{x}_{\rm new},\label{Equ:CMSEnc}\\
 where \quad &
h_c(\bm{x}_{n+1}) = h(z_c) = \Phi(z_c)-z_c\phi(z_c)+\frac{\phi^2(z_c)}{1-\Phi(z_c)}, \quad z_c=\frac{c-\mu_{n+1}}{\sigma_{n+1}}. \notag
\end{align}
Here, $\sigma_{\rm new}^2=\text{\rm Var} [\xi(\bm{x}_{\rm new})|\mathcal{Y}_n]$,  $\rho_{\rm new}(\bm{x}_{n+1}) =\text{\rm Corr} [\xi(\bm{x}_{n+1}),\xi(\bm{x}_{\rm new})|\mathcal{Y}_n]$, $\mu_{n+1}\break =\mathbb{E} [\xi(\bm{x}_{n+1})|\mathcal{Y}_n]$, and $\sigma_{n+1}^2=\text{\rm Var} [\xi(\bm{x}_{n+1})|\mathcal{Y}_n]$ follow from \eqref{Equ:Uncensor_E} and \eqref{Equ:Uncensor_Var}.
$\phi(\cdot)$ and $\Phi(\cdot)$ are the probability density and cumulative distribution functions for the standard normal distribution.
\end{proposition}
\noindent In words, $\mu_{n+1}$ is the predictive mean at $\bm{x}_{n+1}$ given data $\mathcal{Y}_n$, $\sigma_{n+1}^2$ and $\sigma_{\rm new}^2$ are the predictive variances at $\bm{x}_{n+1}$ and $\bm{x}_{\rm new}$, respectively, and $\rho_{\rm new}(\bm{x}_{n+1})$ is the posterior correlation between $\xi(\bm{x}_{n+1})$ and $\xi(\bm{x}_{\rm new})$. Note that the $p$-dimensional integral in \eqref{Equ:CMSEnc} can also be efficiently computed in practice; we provide more discussion later in Corollary \ref{corr:1}. The proof of this proposition can be found in Appendix A.2.

\begin{figure}[!t]
\centering
\includegraphics[width=0.4\textwidth]{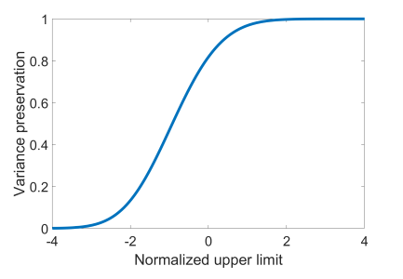}
\caption{\label{Fig:PC} Visualizing the censoring adjustment function $h(z_c)$, where $z_c$ is the normalized right-censoring limit.}
\end{figure}

To glean intuition from the criterion \eqref{Equ:CMSEnc}, we compare it with the existing sequential IMSE criterion \eqref{Equ:Seq_IMSE}. Under no censoring in training data (i.e., $\mathcal{Y}_n = \{\bm{y}_o\}$), \eqref{Equ:Seq_IMSE} can be rewritten as:
\begin{align}
\text{\rm IMSE}(\bm{x}_{n+1}) =\int_{[0,1]^p}\sigma_{\rm new}^2-\rho^2_{\rm new}(\bm{x}_{n+1})\sigma_{\rm new}^2\; d\bm{x}_{\rm new}.
\label{Equ:SeqIMSEnc}
\end{align}   
Comparing \eqref{Equ:SeqIMSEnc} with \eqref{Equ:CMSEnc}, we note a key distinction in the ICMSE criterion: the presence of $h_c(\bm{x}_{n+1})=h(z_c)$, where $z_c$ is the normalized right-censoring limit under the posterior distribution at $\bm{x}_{n+1}$. We call $h(\cdot)$ the \textit{censoring adjustment} function. Fig \ref{Fig:PC} visualizes $h(z_c)$ for different choices of $z_c$. 
Consider first the case of $z_c$ large. From the figure, we see that $h(z_c) \rightarrow 1$ as $z_c \rightarrow \infty$, in which case the proposed ICMSE criterion \eqref{Equ:CMSEnc} reduces to the standard IMSE criterion \eqref{Equ:SeqIMSEnc}. This makes sense intuitively: a large value of $z_c$ (i.e., a high right-censoring limit) means that a new observation at $\bm{x}_{n+1}$ has little posterior probability of being censored at $c$. In this case, the ICMSE criterion (which minimizes predictive variance \textit{under} censoring) should then reduce to the IMSE criterion (which minimizes predictive variance  \textit{ignoring} censoring). 
Consider next the case of $z_c$ small. From the figure, we see that $h(z_c) \rightarrow 0$ as $z_c \rightarrow -\infty$, and the proposed criterion \eqref{Equ:CMSEnc} reduces to the integral of $\sigma_{\rm new}^2$. Again, this makes intuitive sense: a small value of $z_c$ (i.e., a low right-censoring limit) means a new observation at $\bm{x}_{n+1}$ has a high posterior probability of being censored. 
In this case, the ICMSE criterion reduces to the predictive variance of the testing point $\bm{x}_{\rm new}$ given only the first $n$ training data points, meaning a new design point at $\bm{x}_{n+1}$ offers little reduction in predictive variance.
Viewed this way, the proposed ICMSE criterion modifies the standard IMSE criterion by accounting for the
posterior probability of censoring via the censoring adjustment function $h(z_c)$.

Equation \eqref{Equ:CMSEnc} also reveals an important trade-off for the proposed design under censoring. Consider first the standard IMSE criterion \eqref{Equ:SeqIMSEnc}, which minimizes predictive uncertainty under no censoring. Since the first term $\sigma_{\rm new}^2$ does not depend on the new design point $\bm{x}_{n+1}$, this uncertainty minimization is achieved by maximizing the second term $\rho^2_{\rm new}(\bm{x}_{n+1})\sigma_{\rm new}^2$. This can be interpreted as the variance reduction from observing $Y_{n+1}'$ \citep{gramacy2015local}. Consider next the proposed ICMSE criterion \eqref{Equ:CMSEnc}, which maximizes the term $h(z_c)\rho^2_{\rm new}(\bm{x}_{n+1})\sigma_{\rm new}^2$. This can further be broken down into (i) the maximization of variance reduction term $\rho^2_{\rm new}(\bm{x}_{n+1})\sigma_{\rm new}^2$, and (ii) the maximization of the censoring adjustment function $h(z_c)$. Objective (i) is the same as for the standard IMSE criterion -- it minimizes predictive uncertainty assuming no response censoring. Objective (ii), by maximizing the censoring adjustment function $h(z_c)$, aims to minimize the posterior probability of the new design point being censored. Putting both parts together, the ICMSE criterion \eqref{Equ:CMSEnc} features an important trade-off: it aims to find a new design point that jointly minimizes predictive uncertainty (in the absence of censoring) and the posterior probability of being censored.

\subsubsection {Censoring in training data} \label{Sec:Derivation}
We now consider the general case of censored training data $\mathcal{Y}_n=\{\bm{y}_o,\bm{y}_c'\geq \bm{c}\}$. The following proposition gives an explicit expression for the ICMSE criterion.
\begin{proposition} \label{Thm:EMSEwc}
Given the censored data $\mathcal{Y}_n=\{\bm{y}_o,\bm{y}_c'\geq \bm{c}\}$, we have:
\begin{align}
\text{\rm ICMSE}(\bm{x}_{n+1})=\int_{[0,1]^p}\sigma^2_{\rm new}- \boldsymbol{\gamma}_{
n+1,\rm new}^T\bm{\Gamma}_{n+1}^{-1} {\bm{H}}_c(\bm{x}_{n+1}) \bm{\Gamma}_{n+1}^{-1} \boldsymbol{\gamma}_{n+1,
\rm new}\; d\bm{x}_{\rm new},
\label{Equ:generalCMSE}
\end{align}
where $\sigma^2_{\rm new}=\textup{Var} [\xi(\bm{x}_{\rm new})|\mathcal{Y}_n]$, and ${\boldsymbol{\gamma}}_{n+1,\rm new}$ and $\bm{\Gamma}_{n+1}$ follow from \eqref{Equ:Censor_E} and \eqref{Equ:Censor_Var}. The matrix $\bm{H}_c(\bm{x}_{n+1})$ has an easy-to-evaluate expression given in Appendix A.3.
\end{proposition}

\noindent Here, $\sigma^2_{\rm new}$ is the predictive variance at point $\bm{x}_{\rm new}$ conditional on the data $\mathcal{Y}_n$.
The full expression for $(n+1)\times (n+1)$ matrix $\bm{H}_c(\bm{x}_{n+1})$, while easy-to-evaluate, is quite long and cumbersome; this expression is provided in Appendix A.3. The key computation in calculating $\bm{H}_c( \bm{x}_{n+1})$ is evaluating several orthant probabilities from a multivariate normal distribution.  The proof for this proposition can be found in Appendix A.3. Section \ref{Sec:AdaAlgor} and Appendix C provide further details on computation.

While this general ICMSE criterion \eqref{Equ:generalCMSE} is more complex,
its interpretation is quite similar to the earlier criterion  -- its integrand contains a posterior variance term conditional on data $\mathcal{Y}_n$, and a variance reduction term from the potential observation $Y_{n+1}$.
The matrix $\bm{H}_c( \bm{x}_{n+1})$ on the variance reduction term serves a similar purpose to the censoring adjustment function. A large value of $\bm{H}_c( \bm{x}_{n+1})$ (in a matrix sense) suggests a low posterior probability of censoring for a new point $\bm{x}_{n+1}$, whereas a small value suggests a high posterior probability of censoring. This again results in the important trade-off for sequential design under censoring: the proposed ICMSE criterion aims to find the next design point which not only (i) minimizes predictive uncertainty of the fitted model in the absence of censoring, but also (ii) minimizes the posterior probability that the resulting observation is censored. The posterior probability is adaptively learned from the training data, and is not considered by the standard IMSE criterion.

\begin{figure}[!t]
\centering
\includegraphics[width=0.95\textwidth]{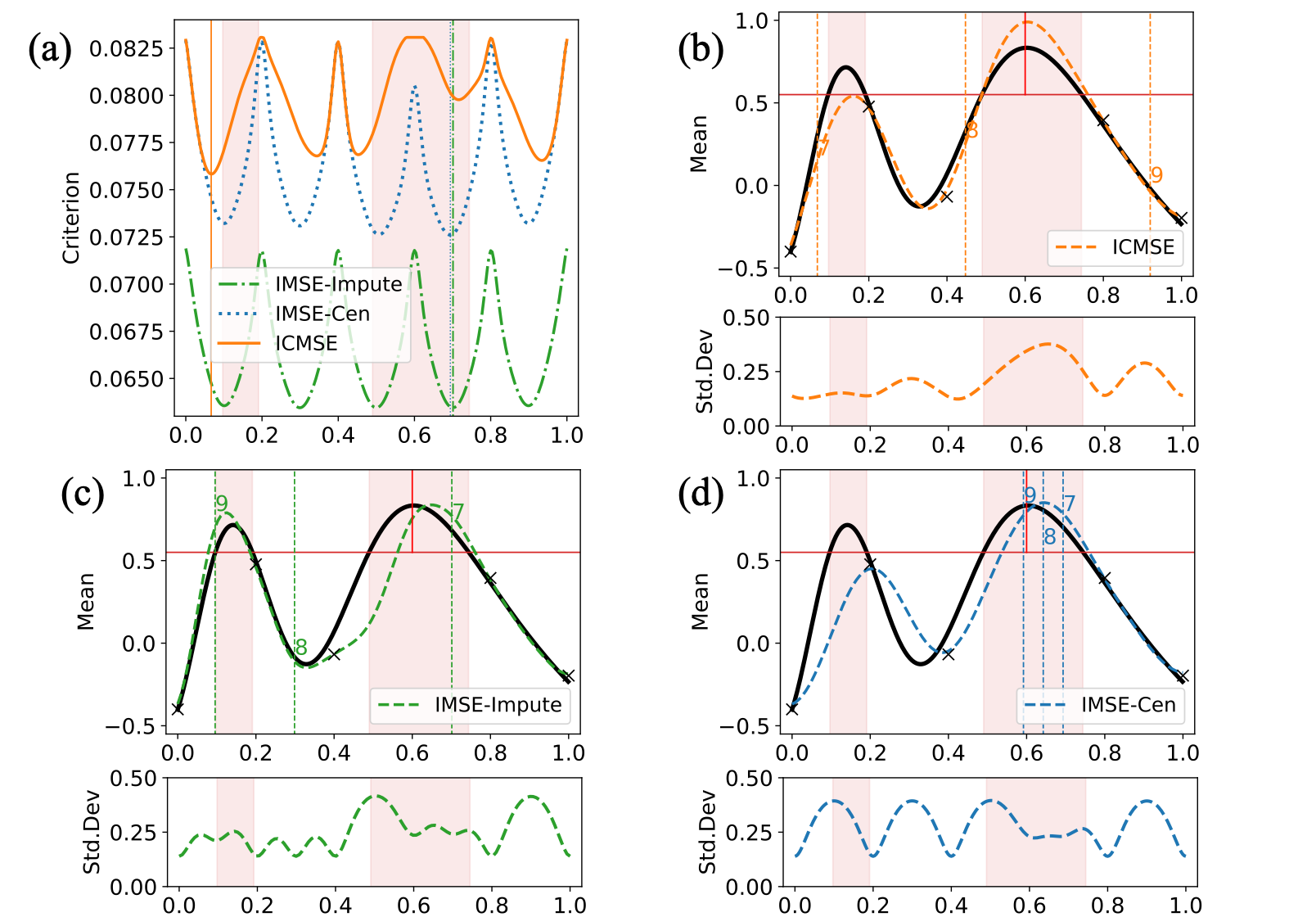}
\caption{ \label{Fig:Illustration} The 1D illustrative example in \eqref{Equ:toyExa1D}: (a) shows the design criteria of the next run $x_7$ for the considered methods.
(b), (c), and (d) show the 3 sequential runs $(x_7^*,x_8^*,x_9^*)$ using ICMSE, IMSE-Impute, and IMSE-Cen, respectively, with the censored regions shaded in red. Top plots show the true function $\xi(\cdot)$ (black line) and the predictor $\hat{\xi}(\cdot)$ (dashed line), with initial observed runs (black crosses, with censored runs indicated by red vertical lines), and sequential points (numbered). Bottom plots show the corresponding predictive standard deviation.}
\end{figure}

\subsection{An illustrative example} \label{Sec:OneRun1D}
We illustrate the ICMSE criterion using a 1D example. Suppose the mean response of the physical experiment is:
\begin{equation}
\xi(x) = 0.5\sin \left(10 (x - 1.02)^2\right)   - 1.25(x - 0.75)(2x-0.25) + 0.2,
\label{Equ:toyExa1D}
\end{equation}
with measurement noise variance $\sigma_\epsilon^2=0.1^2$. Further suppose censoring occurs above an upper limit of $c = 0.55$. The initial design consists of 6 equally-spaced runs, which results in 5 observed runs and 1 censored run.
The Gaussian correlation function is used for $R_{\boldsymbol{\theta}_\xi}$, with model parameters estimated via maximum likelihood.

We compare the ICMSE method \eqref{Equ:generalCMSE} with IMSE methods. Note that, from Section \ref{Sec:ICMSEFormula}, the standard IMSE criterion \eqref{Equ:Seq_IMSE} cannot be directly applied here, since it depends on the potential observation $Y'_{n+1}$ which is unobserved. We adopt the following two variants of IMSE for the censored setting.  The first method, ``IMSE-Impute", is a simple baseline which \textit{imputes} the censored responses in the training data with the known measurement limit $c$. The new design point is then optimized assuming its corresponding response $Y'_{n+1}$ is \textit{not} subject to censoring. The second method, ``IMSE-Cen", integrates the censored runs (in training data) using the \textit{censored} GP model (\ref{Equ:Censor_E}-\ref{Equ:Censor_Var}). The new design point is again optimized assuming its response $Y'_{n+1}$ is \textit{not} subject to censoring. In contrast, the proposed ICMSE method \eqref{Equ:generalCMSE} considers \textit{both} censored training data and the possibility of censoring in the new observation $Y'_{n+1}$ within the design criterion. For a fair comparison, we use the censored GP model (\ref{Equ:Censor_E}-\ref{Equ:Censor_Var}) for evaluating predictive performance for all design methods.

Fig \ref{Fig:Illustration}(a) shows the proposed criterion for the ICMSE method (in orange). It selects the next design point at $x_{7}^*=0.068$, which balances the two desired properties from the ICMSE criterion.
First, it avoids regions with high posterior probabilities of response censoring, due to the presence of $\bm{H}_c(\cdot)$ in \eqref{Equ:generalCMSE}. The next point $x_7^*$, which minimizes \eqref{Equ:generalCMSE}, subsequently \textit{avoids} the censored regions (shaded red), as desired. In contrast, Fig \ref{Fig:Illustration}(a) also shows the design criteria for IMSE-Impute (green) and IMSE-Cen (blue). We see that both IMSE methods choose the next point \textit{within} the censored regions, as the IMSE design criterion does not consider the probability of a new observation being censored. Second, the next point $x_7^*$ chosen by ICMSE minimizes the overall predictive uncertainty for the mean function $\xi(\cdot)$, since the ICMSE criterion is small in regions \textit{away} from existing design points. This can be seen within the region $[0.2,0.5]$, where local minima of the ICMSE criterion are found between training points.

The top plots in Fig \ref{Fig:Illustration}(b)-(d) show the next 3 design points ($x^*_7,x^*_8,x^*_9$) from the 3 considered design methods, as well as the final predictor $\hat{\xi}(\cdot)$ using the censored GP model (\ref{Equ:Censor_E}-\ref{Equ:Censor_Var}) with all 9 points. The bottom plots in Fig \ref{Fig:Illustration}(b)-(d) show the corresponding predictive standard deviation. We see that ICMSE yields noticeably better predictive performance compared to the two IMSE methods.
One reason is that the proposed criterion makes use of the censored GP model for both modeling and design, whereas the two baselines do not.
Table \ref{tab:RMSE1DIll} shows the root mean-squared error (RMSE) after the 3 sequential runs over a test set of 1000 equally-spaced points.
The proposed ICMSE method achieves much smaller errors compared to the two IMSE baselines. We will provide a more comprehensive comparison of predictive performance in Section \ref{Sec:Test1D}.

\begin{table}[!t]
\centering
\begin{tabular}{c |c c c} 
\toprule
& IMSE-Impute & IMSE-Cen & ICMSE\\
 \hline
6 runs & \textbf{0.260} & \textbf{0.260}  & \textbf{0.260} \\
7 runs & 0.214  & 0.214  &\textbf{0.119}\\ 
8 runs & 0.172 & 0.236 & \textbf{0.102}\\
9 runs & 0.153 & 0.203 & \textbf{0.096}\\
\toprule
\end{tabular}
\caption{Predictive performance (in RMSE) for 3 sequential runs in the 1D example \eqref{Equ:toyExa1D}, using ICMSE and the two IMSE baselines (IMSE-Impute and IMSE-Cen).}
\label{tab:RMSE1DIll}
\end{table}

\section{ICMSE design for bi-fidelity modeling} \label{Sec:MultiF}
Next, we extend the ICMSE design to the bi-fidelity setting, where auxiliary computer experiment data are available. We first present the GP framework for bi-fidelity modeling, and extend the earlier ICMSE criterion. We then present an algorithmic framework for efficient implementation, and investigate its performance on two illustrative examples.

\subsection{Modeling framework}
\label{sec:mfmodel}
Let $f(\bm{x})$ denote the \textit{computer} experiment output at input $\bm{x}$. We model $f(\cdot)$ as the GP model:
\begin{equation}
f(\cdot) \sim \text{GP}\{\mu_f,\sigma^2_f R_{\boldsymbol{\theta}_f}(\cdot, \cdot)\}.
\label{Equ:CE}
\end{equation}
Following Section \ref{Sec:PEModel}, let $\xi(\bm{x})$ denote the latent mean response for \textit{physical} experiments  at input $\bm{x}$. We assume that $\xi(\cdot)$ takes the form: 
\begin{align}
\xi(\bm{x}) = f(\bm{x})+\delta(\bm{x}),
\label{Equ:DataFusion}
\end{align}
where $\delta(\bm{x})$ is the so-called \textit{discrepancy} function, quantifying the difference between computer and physical experiments at input $\bm{x}$. Following \cite{KO2001}, we model this discrepancy using a zero-mean GP model:
\begin{equation}
\delta(\cdot) \sim \text{GP}\{ 0, \sigma^2_\delta R_{\boldsymbol{\theta}_\delta}(\cdot,\cdot) \},
\label{Equ:Disc}
\end{equation}
where the prior on $\delta(\cdot)$ is independent of $f(\cdot)$. Here, physical experiments are observed with experimental noise as in Section \ref{Sec:PEModel}, whereas computer experiments are observed without noise.

Suppose $(n-m)$ computer experiments and $m$ physical experiments ($n$ experiments in total) are conducted at inputs $\bm{x}_{1:n}=\{\bm{x}_{1:(n-m)}^f,\bm{x}_{1:m}^\xi\}$, yielding data $\bm{f} = [f_1,\cdots, f_{n-m}]$ and $\mathcal{Y}_{m} = \{\bm{y}_{o}, \bm{y}_c' \geq \bm{c}\}$. Note that censoring occurs only in physical experiments, since computer experiments are conducted via numerical simulations. 
Assuming all model parameters are known (parameter estimation is discussed later in Section \ref{Sec:AdaAlgor}), the mean response $\xi(\bm{x}_{\rm new})$ at a new input $\bm{x}_{\rm new}$ has the following conditional mean and variance:
\begin{align}
\hat{\xi}(\bm{x}_{\rm new}) = \mathbb{E}[\xi(\bm{x}_{\rm new})|\bm{f},\mathcal{Y}_{m}]=\mu_f + \boldsymbol{\gamma}_{n,\rm new}^T\bm{\Gamma}_n^{-1}\left([\bm{f}, \bm{y}_{o}, \hat{\bm{y}}_{c}]^T - \mu_f \textbf{1}_{n} \right), \label{Equ:Fusion_E} \\
s^2(\bm{x}_{\rm new}) = \text{Var}[\xi(\bm{x}_{\rm new})|\bm{f},\mathcal{Y}_{m}]=\sigma^2_f+\sigma^2_\delta-\boldsymbol{\gamma}_{n,\rm new}^T(\bm{\Gamma}_n^{-1}-\bm{\Gamma}_n^{-1}{\bm{\Sigma}}\bm{\Gamma}_n^{-1})\boldsymbol{\gamma}_{n,\rm new},
\label{Equ:Fusion_Var}
\end{align}  
where $\boldsymbol{\gamma}_{n,\rm new}=\sigma^2_f[R_{\boldsymbol{\theta}_f}(\bm{x}_i, \bm{x}_{\rm new})]_{i=1}^{n}+\sigma^2_\delta[\bm{0}_{n-m},R_{\boldsymbol{\theta}_\delta}(\bm{x}_i, \bm{x}_{\rm new})]_{i=1}^{m}$ is the covariance vector, and $\bm{\Gamma}_n=\sigma^2_f{[R_{\boldsymbol{\theta}_f}(\bm{x}_i, \bm{x}_j)]_{i=1}^{n}}_{j=1}^{n} + \text{diag}\big( \bm{0}_{n-m},\sigma^2_\epsilon \bm{I}_{m}+\sigma^2_\delta \times \break {[R_{\boldsymbol{\theta}_\delta}(\bm{x}_i, \bm{x}_j)]_{i=1}^m}_{j=1}^m \big)$ is the covariance matrix.
Here, $\hat{\bm{y}}_{c}=\mathbb{E}[\bm{y}_{c}'|\bm{f},\bm{y}_{o},\bm{y}_{c}'\geq \bm{c}]$ is the expected response for latent vector $\bm{y}_{c}'$ given data $\{\bf{f},\mathcal{Y}_m \}$, and ${\bm{\Sigma}}_{c}=\text{Var}[\bm{y}_{c}'|\bm{f},\bm{y}_{o},\bm{y}_{c}'\geq \bm{c}]$ is its conditional variance, with ${\bm{\Sigma}} = \text{diag}(\textbf{0}_{n-n_c},{\bm{\Sigma}}_{c})$. 
While such equations appear quite involved, they are simply the bi-fidelity extensions of the earlier GP modeling equations \eqref{Equ:Censor_E} and \eqref{Equ:Censor_Var}. For simplicity, we have overloaded some notations from \eqref{Equ:Censor_E} and \eqref{Equ:Censor_Var} here; the difference should be clear from the context.

\subsection{Bi-fidelity design criterion}
Now, we extend the ICMSE design to the bi-fidelity setting. The goal is to design \textit{physical} experiment runs (which may be censored), given auxiliary computer experiment data (which are not censored).

Under the above bi-fidelity GP model, the following proposition gives an explicit expression for the ICMSE design criterion.
\begin{proposition}\label{Thm:MFICMSE}
With experimental data $\{\bm{f},\mathcal{Y}_{m}\}$, the proposed 
ICMSE criterion has the following explicit expression:
\begin{align}
\text{\rm ICMSE}(\bm{x}_{n+1})=&\int_{[0,1]^p} \mathbb{E}_{Y_{n+1}|\bm{f},\mathcal{Y}_m}\left[\text{\rm Var}(\xi(\bm{x}_{\rm new})|\bm{f},\mathcal{Y}_m,Y_{n+1}) \right] \; d\bm{x}_{\rm new} \notag \\
=& \int_{[0,1]^p} \sigma^2_{\rm new}- {\boldsymbol{\gamma}}_{n+1,
\rm new}^T\bm{\Gamma}_{n+1}^{-1} {\bm{H}}_c(\bm{x}_{n+1}) \bm{\Gamma}_{n+1}^{-1} \boldsymbol{\gamma}_{n+1,
\rm new} \; d\bm{x_{\rm new}}, 
\label{Equ:generalCMSEDF}
\end{align}
where $\sigma^2_{\rm new}=\textup{Var} [\xi(\bm{x}_{\rm new})|\bm{f},\mathcal{Y}_m]$, and ${\boldsymbol{\gamma}}_{n+1,\rm new}$ and $\bm{\Gamma}_{n+1}$ follow from \eqref{Equ:Fusion_E} and \eqref{Equ:Fusion_Var}. The matrix $\bm{H}_c(\bm{x}_{n+1})$ has an easy-to-evaluate expression given in Appendix B.1.
\end{proposition}

\noindent 
The proof can be found in Appendix B.1. The following corollary gives a simplification of \eqref{Equ:generalCMSEDF} under a product correlation structure.

\begin{corollary}
\label{corr:1}
Suppose $R_{\boldsymbol{\theta}_f}(\cdot,\cdot)$ and $R_{\boldsymbol{\theta}_\delta}(\cdot,\cdot)$ are product correlation functions: 
\begin{equation}
R_{\boldsymbol{\theta}_f} (\bm{x},\bm{x}')=\prod_{l=1}^p R_{\boldsymbol{\theta}_f}^{(l)} (x_l,x_l'), \quad R_{\boldsymbol{\theta}_\delta} (\bm{x},\bm{x}')=\prod_{l=1}^p R_{\boldsymbol{\theta}_\delta}^{(l)} (x_l,x_l'),
\label{Equ:prodcorr}
\end{equation}
with $\bm{x} = [x_{1}, \cdots, x_{p}]^T$.
Then, the ICMSE criterion \eqref{Equ:generalCMSEDF} can be further simplified as:
\begin{align}
\text{\rm ICMSE}(\bm{x}_{n+1})= \bar{\sigma}^2- \text{\rm tr}\left(\bm{\Gamma}_{n+1}^{-1}  \bm{H}_c(\bm{x}_{n+1}) \bm{\Gamma}_{n+1}^{-1} \boldsymbol{\Lambda} \right), 
\label{Equ:closedformCMSEDF}
\end{align}
where $\bar{\sigma}^2 = \int \sigma^2_{\rm new} \;d \bm{x}_{\rm new}$, 
and $\boldsymbol{\Lambda}$ is an $(n+1) \times (n+1)$ matrix with $(i,j)^{th}$ entry:
\begin{align}
\begin{split}
&\Lambda_{ij} = \prod_{l=1}^p \left[\int_0^1 
\zeta^{(l)}(x_{i,l},x) \zeta^{(l)}(x_{j,l},x)\; dx \right], \quad \text{\rm and }  \\ &\zeta^{(l)}(z,x)  = R_{\boldsymbol{\theta}_f}^{(l)} (z,x) +\mathds{1}_{\{i> (n-m)\}}R_{\boldsymbol{\theta}_\delta}^{(l)} (z,x).
\end{split}
\label{Equ: EntryLambda}
\end{align}
\end{corollary}
\noindent The key simplification from Corollary \ref{corr:1} is that it reduces the $p$-dimensional integral in the ICMSE criterion \eqref{Equ:generalCMSEDF} to a product of 1D integrals, which are more easily computed. Furthermore, if Gaussian correlation functions are used, these integrals can be reduced to error functions, which yield an easy-to-evaluate design criterion for ICMSE (see Appendix B.2 for details). Given the computational complexities of censored data, this simplification allows for efficient design optimization. Corollary \ref{corr:1} is motivated by the simplification of the IMSE criterion in \cite{sacks1989designs}. The proof can be found in Appendix B.2. 

The interpretation of the bi-fidelity ICMSE criterion \eqref{Equ:generalCMSEDF} is analogous to that of the single-fidelity ICMSE criterion \eqref{Equ:generalCMSE}. Similar to the censoring adjustment function, the matrix $\bm{H}_c(\cdot)$ factors in the posterior probability of censoring over the input space, and is used to adjust the variance reduction term in the criterion. Viewed this way, the ICMSE criterion \eqref{Equ:generalCMSEDF} provides the same design trade-off as before: the next design point should jointly (i) avoid censored regions by adaptively identifying such regions from data at hand, and (ii) minimize predictive uncertainty from the GP model.

\subsection{An adaptive algorithm for sequential design} \label{Sec:AdaAlgor} 

We present next an adaptive algorithm \texttt{ICMSE} for implementing the proposed ICMSE design. This algorithm applies for both the single-fidelity setting (with flag $I_{\rm BF}=0$) in Section \ref{Sec:SingeF} and the bi-fidelity setting (with flag $I_{\rm BF}=1$) in Section \ref{Sec:MultiF}.
First, an initial $n_{\rm ini}$-point design is set up for initial experimentation: physical experiments for the single-fidelity setting, and computer experiments for the bi-fidelity setting. In our implementation, we used the maximum projection (MaxPro) design proposed by \cite{joseph2015maximum}, which provides good projection properties and thereby good GP predictive performance. Next, the following two steps are performed iteratively: (i) using observed data $\{\bm{f},\mathcal{Y}_{m}\}$, the GP model parameters are estimated using maximum likelihood, (ii) the next design point $\bm{x}_{n+1}^*$ is then obtained by minimizing the ICMSE criterion (equation \eqref{Equ:generalCMSE} for the single-fidelity setting, equation \eqref{Equ:generalCMSEDF} for the bi-fidelity setting), along with its corresponding response $Y_{n+1}$. This is then repeated until a desired number of samples is obtained.

\begin{algorithm} [!t]
    \begin{algorithmic}[1]
\If{$I_{\rm BF} = 0$}  \Comment{Single-fidelity}
\State Generate an $n_{\rm ini}$-run initial MaxPro design $\bm{x}_{1:n_{\rm ini}}$
\State Collect initial data $\mathcal{Y}_{n_{\rm ini}}$ at inputs $\bm{x}_{1:n_{\rm ini}}$ from physical experiments
\State Estimate model parameters $\{\mu_\xi, \sigma^2_\xi, \boldsymbol{\theta}_\xi\}$ using MLE from initial data $\mathcal{Y}_{n_{\rm ini}}$
\Else  \Comment{Bi-fidelity}
\State Generate an $n_{\rm ini}$-run initial MaxPro design $\bm{x}_{1:n_{\rm ini}}$
\State Collect initial data $\bm{f}$ at inputs $\bm{x}_{1:n_{\rm ini}}$ from computer experiments
\State Estimate model parameters $\{\mu_f,\sigma^2_f, \boldsymbol{\theta}_f\}$ using MLE from $\mathcal{Y}_{n_{\rm ini}}$, and let $\sigma^2_\delta=0$
\EndIf
\For{$k = n_{\rm ini}+1 , \cdots , n_{\rm ini} + n_{\rm seq}$}   \Comment{$n_{\rm seq}$ sequential runs}
\If{$I_{\rm BF} = 0$}  
\State  Obtain new design point $\bm{x}_{k}^*$ by minimizing ICMSE criterion  \eqref{Equ:generalCMSE}
\Else  
\State Obtain new design point $\bm{x}_{k}^*$ by minimizing ICMSE criterion \eqref{Equ:generalCMSEDF}
\EndIf
\State Perform experiment at $\bm{x}_k^*$  and collect response $Y_{k}$ (which may be censored)
\State Update model parameter estimates using new data
\EndFor
\end{algorithmic}
    \caption{\texttt{ICMSE}($n_{\rm ini}$, $n_{\rm seq}$, $c$, $I_{\rm BF}$): Adaptive design under censoring
    }
    \label{Alg:ICMSE}
\end{algorithm}

To optimize the ICMSE criterion,
we use standard numerical optimization methods in the \textsf{R} package \texttt{nloptr} \citep{ypma2014nloptr}, in particular, the Nelder-Mead method \citep{nelder1965simplex}. 
The main computational bottleneck in optimization is evaluating moments of the truncated multivariate normal distribution for $\bm{H}_c(\cdot)$ (see equations (A.10) and (B.3) in the supplementary article, Chen et al., 2021). In our implementation, these moments are efficiently computed using the \textsf{R} package \texttt{tmvtnorm} \citep{wilhelm2010tmvtnorm}. Appendix C details further computational steps for speeding-up design optimization, involving an approximation of the expected variance term via a plug-in estimator. Similar to the standard IMSE criterion, the ICMSE criterion can be quite multi-modal. We therefore suggest performing multiple random restarts of the optimization, and taking the solution with the best objective value as the new design point.

\subsection{Illustrative examples with adaptive design algorithm} \label{Sec:Test1D}

We first illustrate the proposed algorithm \texttt{ICMSE} on a 1D bi-fidelity  example. Suppose the computer simulation is given by
\begin{align}
 f(x) &= 0.5\sin \left(10 (x - 1.02)^2\right)  + 0.1,
 \label{eq:1Dbiexp}
\end{align}
with the same physical experiment settings as in Section \ref{Sec:OneRun1D}. We begin with an $n_{\rm ini}=6$-run equally-spaced points $x^f_{1:6}=\{(i-1)/5\}_{i=1}^6$ for computer experiments. We then perform a sequential $n_{\rm seq}=20$-run design for physical experiments using the algorithm \texttt{ICMSE}. The Gaussian correlation function is used for both GPs. In addition to the two IMSE methods in Section \ref{Sec:OneRun1D}, we consider an additional ``sequential MaxPro'' method \citep{joseph2016rejoinder}, which implements a sequential space-filling design. Again, for fair comparison, we use the censored GP model (\ref{Equ:Censor_E}-\ref{Equ:Censor_Var}) for evaluating predictions for all design methods. The simulation is replicated 20 times.

\begin{figure}[!t]
\centering
\includegraphics[width=0.99\textwidth]{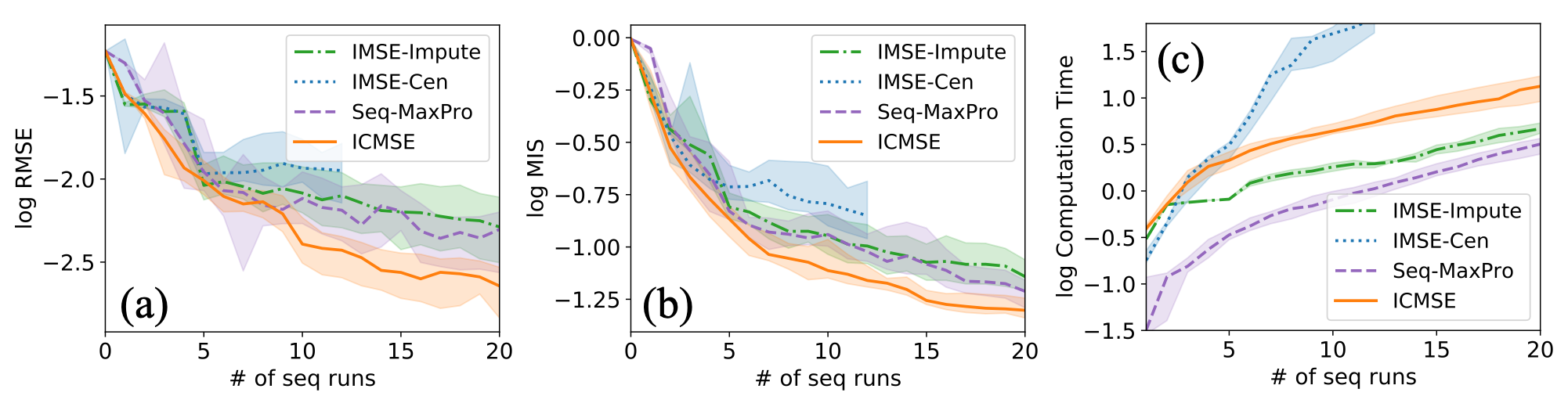}
\caption{\label{Fig:Test1D} The 1D bi-fidelity example in \eqref{eq:1Dbiexp}: (a), (b), and (c) show the log-RMSE, log-MIS, and log-computation time (in seconds) over the number of sequential runs for each method.
Solid lines mark the median over the 20 replications, and shaded regions mark the 25\%-75\% quantiles.}

\end{figure}

We consider two evaluation metrics for predictive performance: RMSE and the interval score proposed in \cite{gneiting2007strictly}. The first assesses predictive accuracy, and the second assesses uncertainty quantification. The $(1-\alpha)$\% interval score is defined as
\begin{align}
    \text{IS}({\xi}_l,{\xi}_u;\xi) =  ({\xi}_u-{\xi}_l)+\frac{2}{\alpha}({\xi}_l-\xi)_++\frac{2}{\alpha}(\xi-{\xi}_u)_+ ,
\end{align}
where $(a)_+ = \max(a,0)$, $\xi$ is the ground truth, and $[{\xi}_l,{\xi}_u] $ is an $(1-\alpha)$\% predictive interval. Here, we set $1-\alpha = 68\%$, with predictive interval $[\hat{\xi}-\sqrt{s^2},\hat{\xi}+\sqrt{s^2}]$, where $\hat{\xi}$ and $s^2$ are obtained from \eqref{Equ:Fusion_E} and \eqref{Equ:Fusion_Var}. The mean interval score (MIS) is then computed over the entire test set. We also compared computation time on a 1.4 GHz Quad-Core Intel Core i5 laptop.

Fig \ref{Fig:Test1D} shows the log-RMSE (a), log-MIS (b), and log-computation time (c) for the 4 considered methods. The ICMSE method yields noticeable improvements over the IMSE and sequential MaxPro methods, with smaller RMSE and MIS values for most sequential run sizes.
One reason for this is that the proposed method integrates the possibility of a new observation being censored directly within the design criterion, which allows it to minimize predictive uncertainty by avoiding censored regions.
While ICMSE requires more computation time compared to the two baseline methods, the computation complexities appear to be comparable. Here, the IMSE-Cen method is terminated early after 12 sequential runs, due to numerical instabilities (and thereby expensive computation) in evaluating the predictive equations. This is because, by ignoring censoring, IMSE-Cen overestimates the potential variance reduction in censored regions, leading to many sequential points very close together in such regions.

\begin{figure}[!t]
\centering
\includegraphics[width=0.8\textwidth]{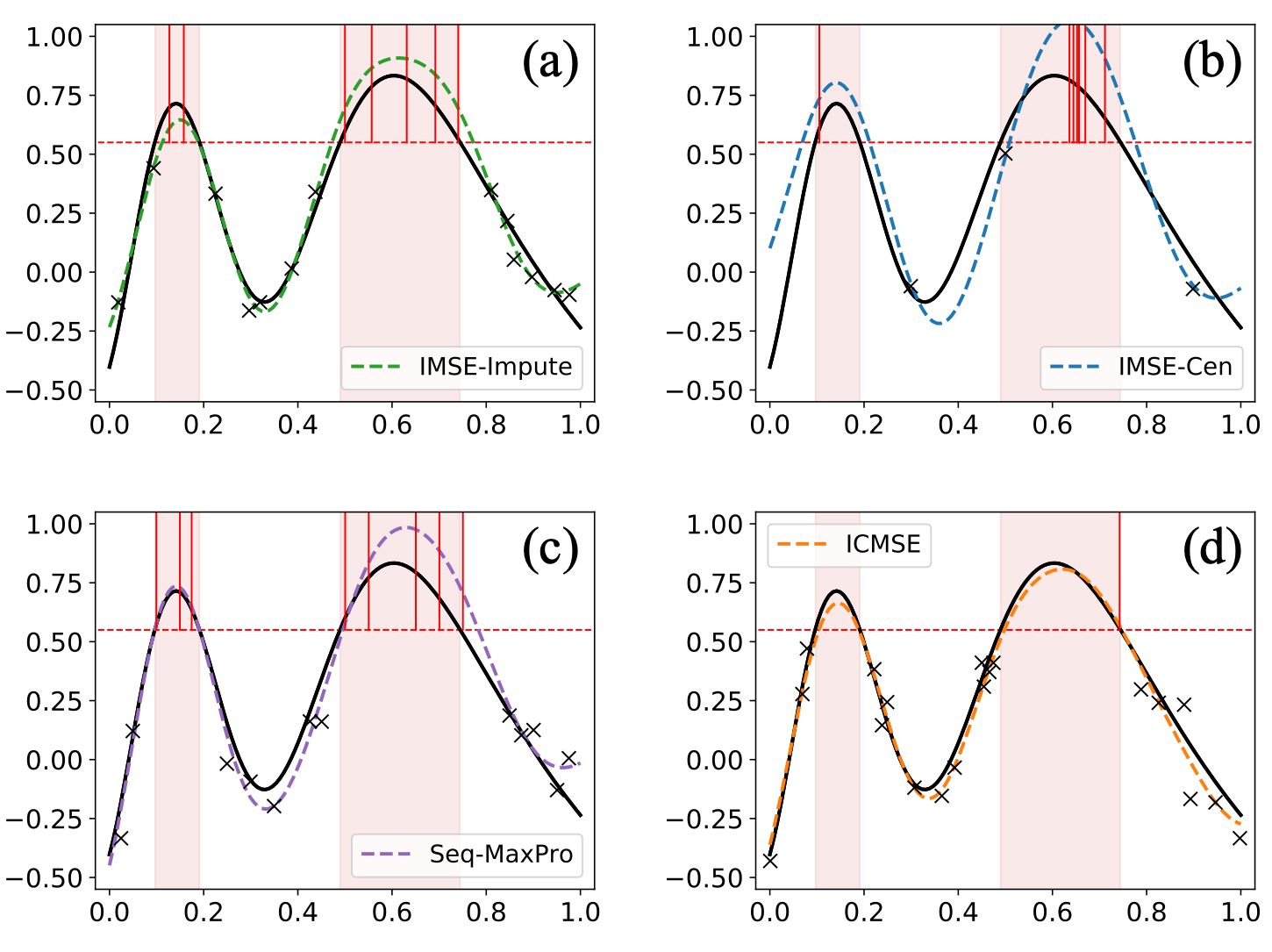}
\caption{\label{Fig:Test1Dallmethod} One replication in the 1D bi-fidelity example \eqref{eq:1Dbiexp}: (a)-(d) show the observed runs (black crosses, with censored runs indicated by red vertical lines) and the predictor $\hat{\xi}(\cdot)$ (dotted lines), using the 4 considered methods. Here, black lines mark the mean physical experiment $\xi(\cdot)$, and shaded regions mark the censored regions. }

\end{figure}

Fig \ref{Fig:Test1Dallmethod} shows the sequential design points and the predicted mean responses $\hat{\xi}(\cdot)$ for a single replication. Compared to existing methods, ICMSE yields visually improved prediction in both the censored (shaded) and uncensored (clear)  regions.
One reason for this is that the ICMSE criterion chooses points which jointly (i) avoid censored regions and (ii) minimize predictive uncertainty. For (i), note that only 1/20 = 5\% of sequential runs are censored for ICMSE, whereas 7/20 = 35\%, 8/20 = 40\%, and 9/12 = 75\% of sequential runs are censored for IMSE-Impute, MaxPro, and IMSE-Cen, respectively. This shows that ICMSE effectively estimates the posterior probability of censoring, and avoids regions with high probabilities for sampling. For (ii), Fig \ref{Fig:Test1Dallmethod}(d) shows that the sequential runs from ICMSE are far away from existing points, and also concentrated near the boundary of the censored region. Intuitively, this minimizes predictive uncertainty by ensuring design points well-explore the input space while avoiding losing information due to censoring.

Next, we conduct a 2D simulation. The computer simulation and mean physical experiment functions are taken from \cite{xiong2013sequential}:
\begin{align}
 f(\bm{x}) =& \frac{1}{4}\xi\left(x_1 + \frac{1}{20}, x_2 + \frac{1}{20}\right)
+\frac{1}{4}\xi\left(x_1 +\frac{1}{20},\left(x_2 -\frac{1}{20}\right)_+\right) \label{eq:2Dexp}\\
+& \frac{1}{4}\xi\left(x_1 - \frac{1}{20},x_2 +\frac{1}{20}\right)
+\frac{1}{4}\xi\left(x_1 -\frac{1}{20},\left(x_2 -\frac{1}{20}\right)_+\right), \notag \\
\xi(\bm{x}) =& \left[1-\exp\left(-\frac{1}{2x_2}\right)\right]\frac{2300x_1^3+1900x_1^2+2092x_1+6}{100x_1^3+500x_1^2+4x_1+20},
\end{align}
with measurement variance $\sigma^2_\epsilon=1$, and a right censoring limit of $c=10$. We begin with an initial $n_{\rm ini}=12$-run MaxPro design for the computer experiment, then add $n_{\rm seq}=40$ sequential runs for physical experiments using \texttt{ICMSE}. This is then replicated 20 times.

\begin{table}[t]
\centering
\begin{tabular}{c|ccc|ccc|ccc}
 \toprule
 & \multicolumn{3}{c|}{RMSE}    & \multicolumn{3}{c|}{MIS}    & \multicolumn{3}{c}{Computation Time (in s)} \\
\hline
Sequential runs        &  5 & 15 & 40  &  5 & 15 & 40    &  5 & 15 & 40   \\  \hline\hline
IMSE-Impute &     1.62       &   1.38  &  {1.14}    &       {4.91}  &  {4.03} &  {3.29} &      {4.14}  & {12.18}  & {64.34}\\
IMSE-Cen     &      1.61       & 1.46       & -    &    \textbf{4.57}      &  4.27   &  -     &   25.13 &  121.24  &   -   \\
{Seq-MaxPro} &      1.74       &   1.36   &  1.12    &         5.58  &  4.01  &  3.22 &      \textbf{3.03}   & \textbf{10.18}   & \textbf{57.24} \\
ICMSE   &     \textbf{1.40}      &  \textbf{1.21} & \textbf{0.97}    &     4.58   &  \textbf{3.80}    &  \textbf{3.01}     &  9.77    &   25.01  &   95.67 \\
 \toprule
\end{tabular}
\caption{The median RMSE, MIS, and computation time, under different sequential run sizes for the 3 considered design methods in a 2D bi-fidelity example \eqref{eq:2Dexp}.}
\label{ta:pred_accuracy}
\end{table}

Table \ref{ta:pred_accuracy} summarizes the median RMSE, MIS, and computation time after 5, 15, and 40 sequential runs. We see that ICMSE yields noticeably lower RMSE and MIS, suggesting the proposed design method gives a better predictive performance. While computationally more expensive than IMSE-Impute and MaxPro, the proposed ICMSE method appears more effective at integrating censoring information for sequential design, which leads to improved predictive performance.

\section{Case studies} \label{Sec:CS}
We now return to the two motivating applications. For the wafer manufacturing problem (which only has physical experiments), we use the single-fidelity ICMSE method in Section \ref{Sec:SingeF}. For the surgical planning application (which has both computer and physical experiments), we use the bi-fidelity ICMSE method in Section \ref{Sec:MultiF}.

\subsection{Thermal processing in wafer manufacturing} \label{Sec:LHresult}
Consider first the wafer manufacturing application in Section \ref{Sec:MotiExampleLH},
where an engineer is interested in how a wafer chip's heating performance is affected by six process input variables that control wafer thickness, rotation speed, heating laser (i.e., its moving speed, radius, and power), 
and heating time.
The response of interest $\xi(\bm{x})$ is the minimum temperature over the wafer, which provides an indication of the wafer's quality after thermal processing. Standard industrial temperature sensors have a measurement limit of $c=350$\textdegree{}C \citep{wafersensors}, and temperatures greater than this limit are censored in the experiment.

\begin{figure}[!t]
\centering
\includegraphics[width=0.99\textwidth]{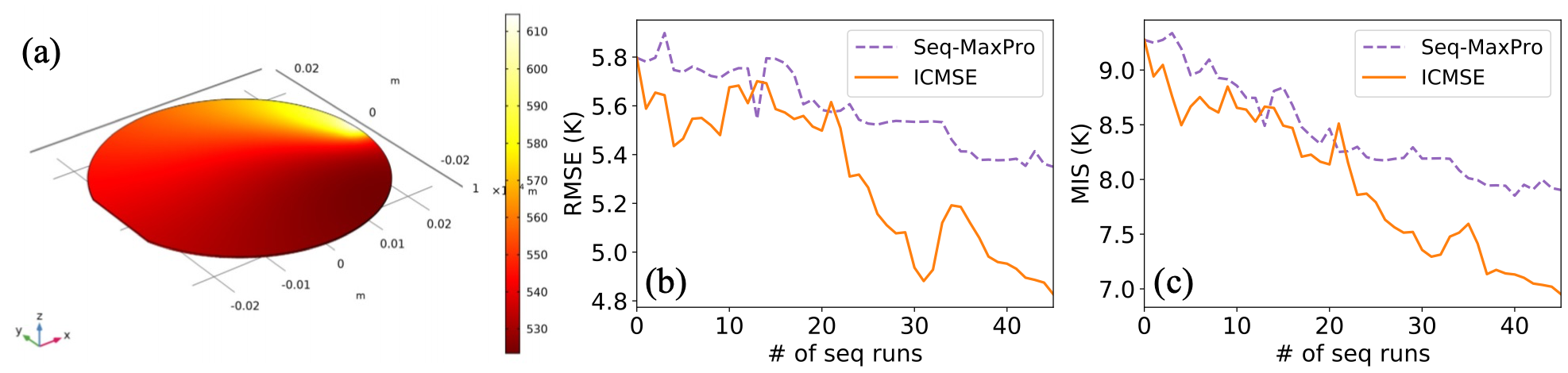}
\caption{\label{Fig:LaserHeatNew} (a) The temperature contour over the wafer chip, simulated using COMSOL Multiphysics. (b) and (c) show the RMSE and MIS of the fitted GP models over the sequential design size, respectively, for the two design methods.}
\end{figure}

As mentioned earlier, certain physical experiments are not only costly (e.g., wafers and laser operation can be expensive), but also time-consuming to perform (e.g., each experiment requires a re-calibration of thermal sensors, as well as a warmup and cooldown of the laser beam). To compare the sequential performance of these methods over a large number of runs, we mimic the costly physical experiments\footnote{The surgical planning application in Section \ref{Sec:MMResult} performs actual physical experiments, but provides fewer sequential runs due to the expensive nature of such experiments.}
with COMSOL Multiphysics simulations (Fig \ref{Fig:LaserHeatNew}(a)), which provides a realistic representation of heat diffusion physics \citep{dickinson2014comsol}. Measurement noise is then added, following an i.i.d. zero-mean normal distribution with standard deviation $\sigma_\epsilon=1.0$\textdegree{}C.

The set-up is as follows. We start with an $n_{\rm ini}=30$-run initial experiment, then perform $n_{\rm seq} = 45$ sequential runs. Note that the total number of $n_{\rm ini} + n_{\rm seq} = 75$ runs is slightly more than the rule-of-thumb sample size of $10p$ recommended by \cite{loeppky2009choosing} -- this is to ensure good predictive accuracy under censoring. Due to the limited budget, the proposed ICMSE method is compared with only the sequential MaxPro method. This is because, from simulations in Section \ref{Sec:Test1D}, it provides the best predictive performance and is the fastest among the three baseline methods. 

The fitted GP models are then tested on temperature data generated (without noise) on a 200-run Sobol' sequence \citep{sobol1967distribution}.
Of these 200 test samples, 25 samples have minimum temperatures which exceed the censoring limit of $c = 350$\textdegree{}C, suggesting that roughly $12.5\%$ of the design space leads to censoring. 
It is important to note that predictive accuracy is desired for \textit{both} censored and uncensored test runs, since the engineering objective is to predict the experimental response surface prior to censoring. This allows industrial engineers to explore a wide range of quality requirements in manufacturing wafers with low temperatures (which are uncensored in experimentation) and high temperatures (which may potentially be censored).

\subsubsection{Predictive performance}
Fig \ref{Fig:LaserHeatNew} compares the RMSE and MIS after $n_{\rm seq} = 45$ sequential runs. While both sequential methods provide relatively steady improvements in RMSE and MIS, the proposed ICMSE method gives a greater predictive improvement over MaxPro. In particular, with 45 sequential runs, ICMSE achieves an RMSE reduction of $(5.8-4.8)/5.8 = 17.2\%$ over the initial 30 runs, which is greater than the RMSE reduction of $(5.8-5.35)/5.8=7.8\%$ for MaxPro. 
Similarly, for MIS, ICMSE achieves a reduction of $(9.28-6.95)/9.28 = 25.1\%$, compared to $(9.28-7.9)/9.28 = 14.8\%$ for MaxPro. 
This can again be explained by the fact that ICMSE jointly avoids censoring and minimizes predictive uncertainty. Here, ICMSE yields no censored measurements, whereas MaxPro yields 5 censored measurements (a censoring rate of $5/45 = 11.1\%$). Moreover, ICMSE \textit{adaptively} chooses points that minimize predictive uncertainty of the GP model under censoring. This can be seen from Fig \ref{Fig:LaserHeatNew}(b) and (c): 
the ICMSE yields progressively lower RMSE and MIS values as sample size increases.

While the ICMSE provides noticeable improvements over the sequential MaxPro, the reductions in RMSE for both methods are only moderate in magnitude. One reason may be that the underlying response surface for minimum temperature is quite non-smooth over the parameter space, which makes it difficult to learn with a limited number of experimental runs, particularly in censored regions.
It is also worth noting, however, that even moderate improvements in predictive accuracy can lead to significant improvements in wafer manufacturing.
As mentioned in Section  \ref{Sec:MotiExampleLH}, the fitted GP model is used to find process settings that jointly minimize operational costs while meeting target quality requirements. The improved predictive model using ICMSE can cut down waste in heating power and reduce the number of wafers to be re-manufactured, which results in significant cost reductions in the wafer manufacturing process.

\subsection{3D-printed aortic valves for surgical planning} \label{Sec:MMResult}
Consider next the surgical planning application in Section \ref{Sec:MotiExampleMM}, which uses state-of-the-art 3D printing technology to mimic biological tissues. Here, doctors are interested in predicting the stiffness of the printed organs with different metamaterial geometries.
We will consider 3 design inputs $\bm{x} = (A,\omega,d)$, which parametrize a standard sinusoidal form of the substructure curve $I(t) = A \sin (\omega t)$,
with diameter $d$ (see Fig \ref{Fig:MMIntro}(b) for a visualization). 
This parametric form has been shown to provide effective tissue-mimicking performance in prior studies \citep{wang2016dual,chen2018efficient}.
The response of interest $\xi(\bm{x})$ is the elastic modulus at a strain level of 8\%, which quantifies the stiffness at a similar load situation inside the human body \citep{wang2016dual}.

We use the bi-fidelity ICMSE design framework in Section \ref{Sec:MultiF}, since a pre-conducted database of computer simulations is available, and we are interested in the sequential design of physical experiments. Computer simulations were performed with finite element analysis \citep{zienkiewicz1977finite} using COMSOL Multiphysics.
Physical experiments were performed in two steps: the aortic valves were first 3D-printed by the Connex 350 machine (Stratasys Ltd.), and then its stiffness was measured by a load cell using uniaxial tensile tests (see Fig \ref{Fig:MMIntro}(c); \citealp{wang2016dual}). 
Here, physical experiments are very costly, requiring expensive material and printing costs, as well as several hours of an experimenter's time per sample.
Censoring is also present in physical experiments; this happens when the force measurement of the load cell exceeds the standard limit of $15N$, corresponding to a modulus upper limit of $c = 0.23 \text{MPa} = 15N \text{(force)}/ 8mm^2 \text{(area)} / 8\% \text{(deformation)}$.

\begin{figure}[t]
\begin{minipage}{0.59\textwidth}
\centering
\includegraphics[width=1\textwidth]{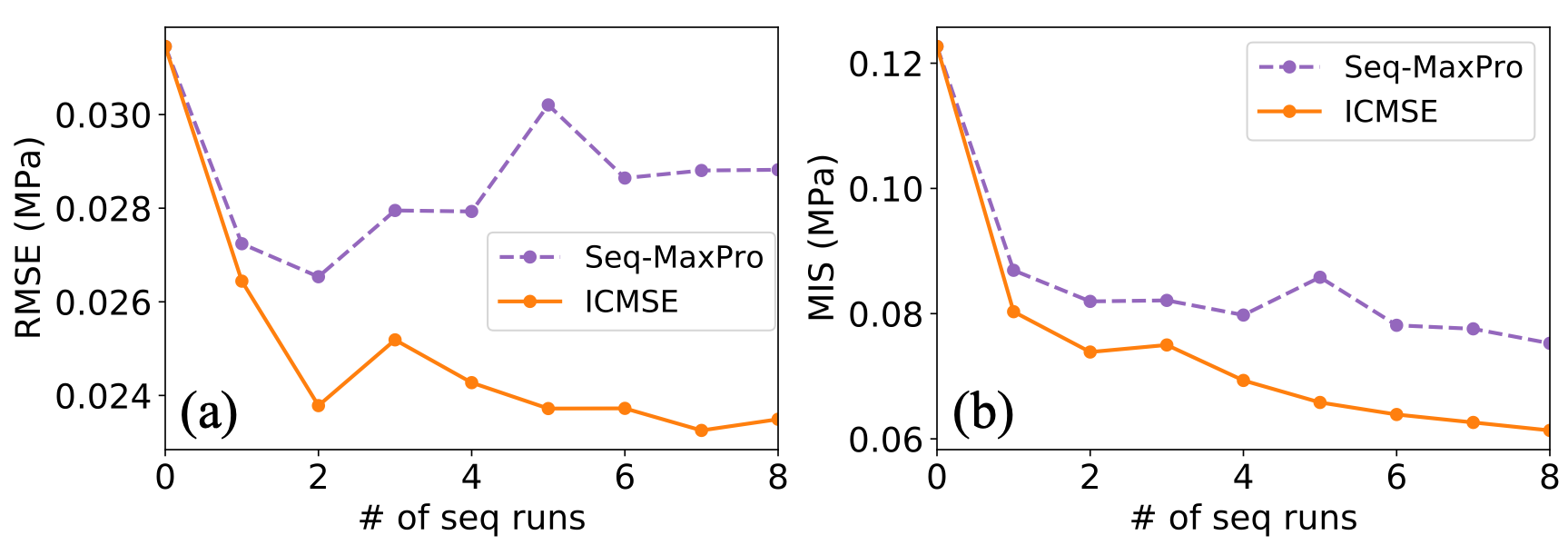}
\captionof{figure}{\label{Fig:MetaMaterialNew} RMSE (a) and MIS (b) for the two sequential design methods, over the number of sequential runs.}
\end{minipage}
\hfill
\begin{minipage}{0.38\textwidth}
\begin{tabular}{c |c c} 
\toprule
& MaxPro & ICMSE\\ 
 \hline
Full & 0.0288 &\textbf{0.0235} \\
Censored & 0.0462   & \textbf{0.0416}   \\ 
Observed & 0.0199 & \textbf{0.0126}  \\
\toprule
\end{tabular}
\captionof{table}{RMSE on the full test set, the 5 censored runs, and the 15 observed runs, for the two sequential design methods.}
\label{Tab:PredSensorNew}
\end{minipage}

\end{figure}

The following design set-up is used. We start with an $n_{\rm ini}=25$-run initial computer experiment design, and then perform $n_{\rm seq}=8$ sequential runs using physical experiments. The limited number of sequential runs is due to the urgent demand of the patients; in such cases, only one to two days of surgical planning can be afforded \citep{chen2018efficient}. Since physical experiments require tedious 3D printing and a tensile test (around 1.5 hours per run), this means only a handful of runs can be performed in urgent cases. 
As before, we compare the proposed ICMSE method with the MaxPro method. The fitted GP models from both methods are tested on the physical experiment data from a 20-run Sobol' sequence.
Among these 20 runs, 5 of them are censored due to the load cell limit; in such cases, we re-perform the experiment using a different testing machine with a wider measurement range. The re-experimentation is typically \textit{not} feasible in urgent surgical scenarios, since it requires even more time-consuming tests and higher material costs.

\subsubsection{Predictive performance}

Fig \ref{Fig:MetaMaterialNew} compares the predictive performance of the two design methods over $n_{\rm seq} = 8$ sequential runs. 
While MaxPro shows
some stagnation in RMSE and MIS improvement, ICMSE yields more noticeable improvements as sample size increases.
More specifically, ICMSE achieves an RMSE reduction of roughly $(0.0315-0.0235)/0.0315=25.4\%$ over the initial GP model (fitted using 25 computer experiment runs), which is much greater than the RMSE reduction of $(0.0315-0.0288)/0.0315=8.57\%$ for MaxPro. Similar improvements can be seen by inspecting MIS. This can again be attributed to the key design trade-off. ICMSE adaptively identifies and avoids censored regions on the design space using the fitted bi-fidelity  model \eqref{Equ:Fusion_E}. Here, the proposed method yields no censored measurements, whereas MaxPro yields 3 censored measurements (a censoring rate of $3/8=37.5\%$).
Furthermore, in contrast to MaxPro, which encourages physical runs to be ``space-filling'' to the initial computer experiment runs, ICMSE instead incorporates censoring information within an adaptive design scheme, which allows for improved predictive performance.

We investigate next the predictive performance of both designs within the \textit{censored} region. This region (corresponding to stiff valves) is important for prediction, since such valves can be used to mimic older patients \citep{sicard2018aging}. 
We divide the test set (20 runs in total) into two categories: observed runs (15 in total) and censored runs (5 in total). The responses for the latter are obtained via new experiments on a stiffer load cell (which, as mentioned in Section \ref{Sec:MotiExampleMM}, is typically not feasible in practice).
Table \ref{Tab:PredSensorNew} compares the RMSE of the two methods for the censored and uncensored test runs. 
For both methods, the RMSE for observed test runs is much smaller than that for censored test runs, which is as expected.
For censored test runs, ICMSE also performs slightly better than MaxPro, with $(0.0462-0.0416)/0.0462=9.9\%$ lower RMSE.
One reason for this is that ICMSE encourages new runs near (but not within) censored regions (see Fig \ref{Fig:Test1D}), to maximize information under censoring. Because of this adaptivity, ICMSE achieves better predictive performance within the censored region, without putting any sequential runs in this region.

The improved performance for ICMSE can greatly improve the surgery planning procedure.
As discussed in Section \ref{Sec:MotiExampleMM}, 
the fitted GP model is used to optimize a polymer geometry which mimics a patient's tissue stiffness. An improved predictive model leads to better tissue-mimicking performance of the printed valves, which then translates to improved surgery success rates. Indeed, in a recent study \cite{chen2018JASA}, it was shown that a predictive model with 42\% improvement in predictive performance leads to six-fold error reduction for tissue-mimicking. We would expect a similar improvement in tissue-mimicking performance here, when comparing ICMSE with the baseline design methods. The resulting improved artificial valves from ICMSE then leads to improved success rates for heart surgeries.

\subsubsection{Discrepancy modeling}
The ICMSE method can also yield valuable insights on the discrepancy between computer simulation and reality. The learning of this discrepancy from data is important for several reasons: it allows doctors to (i) pinpoint where simulations may be unreliable, (ii) identify potential root causes for this discrepancy, and (iii) improve the simulation model to better mimic reality. In our modeling framework, this discrepancy can be estimated as:
\begin{equation}
\hat{\delta}(x) = \hat{\xi}(x) - \hat{f}(x),
\end{equation}
where $\hat{\xi}(x)$ is the predictor for the physical experiment mean, fitted using 25 initial computer experiment runs and 8 physical experiment runs, and $\hat{f}(x)$ is the computer experiment model fitted using only the 25 initial runs.

\begin{figure}[!t]
\centering
\includegraphics[width=0.99\textwidth]{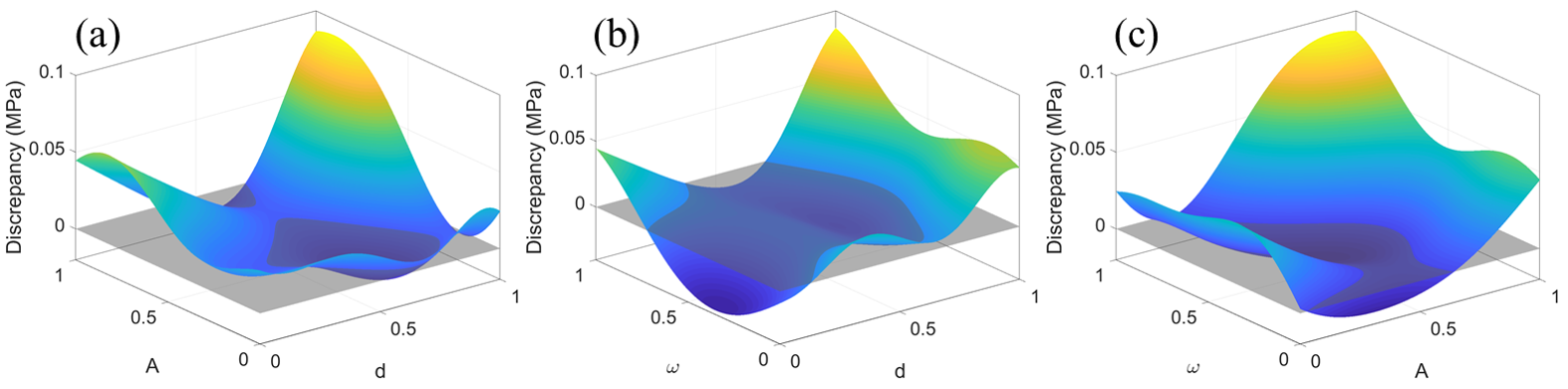}
\caption{\label{Fig:Discrep} Visualization of the estimated discrepancy $\hat{\delta}(\cdot)$ (a) over $d$ and $A$, with fixed $\omega = 1$, (b) over $d$ and $\omega$, with fixed $A = 1$, and (c) over $A$ and $\omega$, with fixed $d = 1$.}
\end{figure}

Fig \ref{Fig:Discrep} shows the fitted discrepancy $\hat{\delta}(x)$ as a function of each pair of design inputs, with the third input fixed.
These plots reveal several interesting insights.
First, when the diameter $d$ is moderate (i.e., $d \in [0.2,0.7]$), Fig \ref{Fig:Discrep}(a) and (b) show that the discrepancy is quite small; however, when $d$ is small (i.e., $[0,0.2]$) or large (i.e., $[0.7,1]$), the discrepancy can be quite large. 
This is related to the limitations of finite element modeling.
When diameter $d$ is small, the simulations can be inaccurate, since the mesh size would be relatively large compared to $d$.  
When diameter $d$ is large, simulations can again be inaccurate, due to the violation of the perfect interface assumption between the two printed polymers.
Second, from Fig \ref{Fig:Discrep}, model discrepancy also appears to be largest when all design inputs are large (i.e., close to 1). This suggests that simulations can be unreliable, when the stiff material is both thick ($d \approx 1$) and fluctuating ($\omega \approx 1, A \approx 1$).
Finally, the model discrepancy is mostly positive over the design domain, revealing smaller stiffness evaluation via simulation compared to physical evaluation.
This may be caused by the hardening of 3D-printed samples due to exposure to natural light, as an aging property for the polymer family (e.g., see \citealp{liao1998long}).
Therefore, the printed aortic valves should be stored in dark storage cells for surgical planning to minimize exposure to light.

\section{Conclusion}
\label{sec:conclusion}
In this paper, we proposed a novel integrated censored mean-squared error (ICMSE) method for adaptively designing physical experiments under response censoring. The ICMSE method iteratively performs two steps: it first estimates the posterior probability of a new observation being censored, and then selects the next design point which yields the greatest reduction in predictive uncertainty under censoring.
We derived easy-to-evaluate expressions for the ICMSE design criterion in both the single-fidelity and bi-fidelity settings, and presented an adaptive design for efficient implementation. We then demonstrated the effectiveness of the proposed ICMSE method over existing methods in real-world applications on 3D-printed aortic valves for surgical planning and thermal processing in wafer manufacturing. 
An \textsf{R} package is currently in development and will be released soon.

Looking ahead, there are several interesting directions to be explored. In this work, the censoring limit $c$ is assumed to be known. While this is true for the two motivating applications, there are other problems where $c$ is unknown and needs to be learned from data; it would be useful to extend ICMSE for such problems. 
Another direction is to explore the connection between the ICMSE method and the multi-points expected improvement method \citep{ginsbourger2010kriging}, which may speed up design optimization via rejection sampling.
Finally, for the bi-fidelity ICMSE, it would be interesting to explore more elaborate design schemes that allow for additional computer experiments to be added sequentially.



\appendix

\section{Single-fidelity ICMSE design criterion}

\subsection{A useful intermediate derivation}

We present first a simplified expression for the design criterion (2.7), which will aid in later derivations.

Let $Y_{n+1}'$ be the latent response at $\bm{x}_{n+1}$ \textit{prior} to censoring, and $Y_{n+1}=Y_{n+1}'(1-\mathcal{I}(\bm{x}_{n+1}))+c\;\mathcal{I}(\bm{x}_{n+1})$ be the corresponding response \textit{after} censoring, with $c$ the right-censoring limit. Here, we define the censoring indicator function:
\begin{align*}
\mathcal{I}(\bm{x}_{n+1}) = \mathds{1}_{\{Y_{n+1}'\geq c\}} = \left\{
\begin{matrix}
0 & \text{ if } & Y_{n+1}'\geq c  \\ 
1 & \text{ if } & Y_{n+1}'<c
\end{matrix}\right. .
\end{align*}
We can now define the probability of censoring at potential input point $\bm{x}_{n+1}$ as $\lambda=\lambda(\bm{x}_{n+1})=\mathbb{P}[\mathcal{I}(\bm{x}_{n+1})=1|\mathcal{Y}_n]=\mathbb{P}[Y_{n+1}'\geq c|\mathcal{Y}_n]$.

Let $\text{CMSE}(\bm{x}_{n+1},\bm{x}_{\rm new})$ be the integrand of the proposed criterion (2.7). One can decompose this integrand via the total variance formula:
\begin{align*}
   \text{CMSE}(\bm{x}_{n+1},\bm{x}_{\rm new}) =\text{Var} [\xi(\bm{x}_{\rm new})|\mathcal{Y}_n] -   \text{Var}_{Y_{n+1}|\mathcal{Y}_n}\left[ \mathbb{E}\left(\xi(\bm{x}_{\rm new})|Y_{n+1},\mathcal{Y}_n\right)\right].
\end{align*}
Denote the first term $\text{Var} [\xi(\bm{x}_{\rm new})|\mathcal{Y}_n]=\sigma_{\rm new}^2$. As for the second term, we compute variance of the random variable $Z=\mathbb{E}\left(\xi(\bm{x}_{\rm new})|Y_{n+1},\mathcal{Y}_n\right)$ by conditioning on the censoring indicator function $\mathcal{I}=\mathcal{I}(\bm{x}_{n+1})$:
\begin{align*}
    \text{Var}[Z]= \mathbb{E}_{\mathcal{I}}[\text{Var}(Z|\mathcal{I})] + \text{Var}_{\mathcal{I}}[\mathbb{E}(Z|\mathcal{I})].
\end{align*}
Consider first the expected variance term $\mathbb{E}_{\mathcal{I}}[\text{Var}(Z|\mathcal{I})]$. Since $Z$ is a constant when censoring occurs (i.e., $\mathcal{I}=1$), the first term becomes: 
\begin{align}
\label{AppEqu:EV}
\begin{split}
     \mathbb{E}_{\mathcal{I}}[\text{Var}(Z|\mathcal{I})]
     &=(1-\lambda)\text{Var}_{Y_{n+1}|\mathcal{I}=0}[\mathbb{E}(\xi(\bm{x}_{\rm new})|Y_{n+1})]  + \lambda \times 0 \\
     &=(1-\lambda) \left(\text{Var}[\xi(\bm{x}_{\rm new})|Y_{n+1}'\geq c] -
     \mathbb{E}_{Y_{n+1}|Y_{n+1}'<c}[\text{Var}(\xi(\bm{x}_{\rm new})|Y_{n+1})]\right),
\end{split}
\end{align}
where the condition on data $\mathcal{Y}_n$ is omitted for brevity. Consider next the variance of expectation term $\text{Var}_{\mathcal{I}}[\mathbb{E}(Z|\mathcal{I})]$. Note that the random variable $\mathbb{E}[Z|\mathcal{I}]$ follows a two point distribution. Hence, the second term becomes:
\begin{align}
\label{AppEqu:VE}
\begin{split}
     \text{Var}_{\mathcal{I}}[\mathbb{E}(Z|\mathcal{I})]
     &=\lambda(1-\lambda)\left( \mathbb{E}\left[\xi(\bm{x}_{\rm new})|\mathcal{I}=1\right] - \mathbb{E}_{Y_{n+1}|\mathcal{I}=0}\left[\mathbb{E}(\xi(\bm{x}_{\rm new})|Y_{n+1})\right] \right)^2 \\
     &=\lambda(1-\lambda)\left( \mathbb{E}\left[\xi(\bm{x}_{\rm new})|Y_{n+1}'\geq c\right] -  \mathbb{E}\left[\xi(\bm{x}_{\rm new})|Y_{n+1}'< c\right] \right)^2,
\end{split}
\end{align}
where the condition on data $\mathcal{Y}_n$ is again omitted for brevity. Putting these together, we have the following expression for $\text{CMSE} = \text{CMSE}(\bm{x}_{n+1},\bm{x}_{\rm new})$ :
\begin{align}
\label{AppEqu:CMSE}
\begin{split}
     \text{CMSE}
     &=\sigma^2_{\rm new}-\lambda(1-\lambda)\left( \mathbb{E}\left[\xi(\bm{x}_{\rm new})|Y_{n+1}'\geq c\right] -  \mathbb{E}\left[\xi(\bm{x}_{\rm new})|Y_{n+1}'< c\right] \right)^2 \\
     & - (1-\lambda) \left(\text{Var}[\xi(\bm{x}_{\rm new})|Y_{n+1}'\geq c] -
     \mathbb{E}_{Y_{n+1}|Y_{n+1}'<c}[\text{Var}(\xi(\bm{x}_{\rm new})|Y_{n+1})]\right) .     
\end{split}
\end{align}
which again is the integrand of the proposed criterion $\text{ICMSE}(\bm{x}_{n+1})$.

\subsection{Proof of Proposition 1}\label{Sec:AppCMSEnc}

Suppose no censoring in training data, i.e., $\mathcal{Y}_n = \{\bm{y}_o\}$. Using the conditional mean and variance expressions (2.5) and (2.6) for standard GP regression, we have:
\begin{align*}
    \begin{bmatrix}
 Y_{n+1}' \\
   \xi(\bm{x}_{\rm new})
\end{bmatrix} \big| \mathcal{Y}_n\sim \mathcal{N}
\left(
 \begin{bmatrix}
    \mu_{n+1} \\
    \mu_{\rm new}
\end{bmatrix},
 \begin{bmatrix}
    \sigma_{n+1}^2 &  \rho\sigma_{n+1}\sigma_{\rm new} \\
    \rho\sigma_{n+1}\sigma_{\rm new} & \sigma_{\rm new}^2
\end{bmatrix}
\right).
\end{align*}
Here, the predictive means are:
\begin{align*}
\mu_{n+1} &= \mathbb{E}[Y_{n+1}'|\mathcal{Y}_n]= \mu_\xi+\boldsymbol{\gamma}_{n,n+1}^T \bm{\Gamma}_n^{-1}(\bm{y}_o-\mu_\xi \cdot \bm{1}_n), \quad \text{and}\\
\mu_{\rm new} &= \mathbb{E}[\xi(\bm{x}_{\rm new})|\mathcal{Y}_n]= \mu_\xi+\boldsymbol{\gamma}_{n,\rm new}^T \bm{\Gamma}_n^{-1}(\bm{y}_o-\mu_\xi\cdot \bm{1}_n),
\end{align*}
the predictive variances are:
\begin{align*}
\sigma_{n+1}^2 &= \text{Var}[Y_{n+1}'|\mathcal{Y}_n]= \sigma^2_\xi-\boldsymbol{\gamma}_{n,n+1}^T \bm{\Gamma}_n^{-1} \boldsymbol{\gamma}_{n,n+1},\\
\sigma_{\rm new}^2 &= \text{Var}[\xi(\bm{x}_{\rm new}|\mathcal{Y}_n]= \sigma^2_\xi-\boldsymbol{\gamma}_{n,\rm new}^T \bm{\Gamma}_n^{-1} \boldsymbol{\gamma}_{n,\rm new}, \quad \text{and}\\
\rho=\rho_{\rm new}(\bm{x}_{n+1})&=\text{Corr}[Y_{n+1}',\xi(\bm{x}_{\rm new})|\mathcal{Y}_n] = \sigma^2_\xi R_{\boldsymbol{\theta}_\xi}(\bm{x}_{n+1},\bm{x}_{\rm new})-\boldsymbol{\gamma}_{n,n+1}^T \bm{\Gamma}_n^{-1} \boldsymbol{\gamma}_{n,\rm new}.
\end{align*}
Here, $\boldsymbol{\gamma}_{n,n+1} = \sigma^2_\xi\big[R_{\boldsymbol{\theta}_\xi}(\bm{x}_1, \bm{x}_{n+1}), \cdots, R_{\boldsymbol{\theta}_\xi}(\bm{x}_n, \bm{x}_{n+1}) \big]^T$.

We then calculate the first two moments of truncated (bivariant) normal distribution:
\begin{align}
   \mathbb{E}\left[\xi(\bm{x}_{\rm new})|Y_{n+1}'\geq c\right] & = \int _{-\infty} ^{\infty} y_{\rm new} \times \psi_{Y_{\rm new}|Y_{n+1}\geq c}( y_{\rm new})\; d y_{\rm new} \notag \\
    &=\frac{1}{1-\Phi(z_c)}  \int _c^{\infty}  \psi_{Y_{n+1}}( y_{n+1})\;   d y_{n+1} \int _{-\infty} ^{\infty}y_{\rm new}  \psi_{Y_{\rm new}|Y_{n+1}}(y_{\rm new}) \;dy_{\rm new}\notag \\
    &=\mu_{\rm new} + \rho\sigma_{\rm new} \frac{\phi(z_c)}{1-\Phi(z_c)}, \label{AppEqu:NoCensorTerm1}
\end{align}
where $z_c=(c-\mu_{n+1})/\sigma_{n+1}$ is the normalized upper censoring limit, $\psi_X(\cdot)$ is the probability density function (PDF) of random variable $X$, $\phi(\cdot)$ is the PDF of standard normal distribution, and  $\Phi(\cdot)$ is the cumulative distribution function (CDF) of the standard normal distribution. Similarly, we have:
\begin{align}
   \mathbb{E}\left[\xi(\bm{x}_{\rm new})|Y_{n+1}'< c\right]  & =\mu_{\rm new} - \rho\sigma_{\rm new} \frac{\phi(z_c)}{\Phi(z_c)},  \quad \text{and} \label{AppEqu:NoCensorTerm2}\\  
   \text{Var}\left[\xi(\bm{x}_{\rm new})|Y_{n+1}'\geq  c\right] & =\sigma_{\rm new}^2\left[1+\rho^2 z_c\frac{\phi(z_c)}{1-\Phi(z_c)}-\rho^2\left(\frac{\phi(z_c)}{1-\Phi(z_c)}\right)^2\right]. \label{AppEqu:NoCensorTerm3}
\end{align}

Furthermore, since the conditional variance of the joint normal distribution does not depend on the value of $Y_{n+1}$, we have: 
\begin{align}
\mathbb{E}_{Y_{n+1}|Y_{n+1}'<c}[\text{Var}(\xi(\bm{x}_{\rm new})|Y_{n+1})] = 
\text{Var}(\xi(\bm{x}_{\rm new})|Y_{n+1}) = (1-\rho^2)\sigma_{\rm new}^2.
\label{AppEqu:NoCensorTerm4}
\end{align}
Finally, plugging \eqref{AppEqu:NoCensorTerm1}, \eqref{AppEqu:NoCensorTerm2}, \eqref{AppEqu:NoCensorTerm3}, and \eqref{AppEqu:NoCensorTerm4} back to \eqref{AppEqu:CMSE}, and using the fact that the probability of censoring $\lambda=\mathbb{P}[Y_{n+1}'\geq c|\mathcal{Y}_n]=1-\Phi(z_c)$, we have:
\begin{align*}
   \text{CMSE}(\bm{x}_{n+1},\bm{x}_{\rm new})= \sigma_{\rm new}^2-\sigma_{\rm new}^2\rho^2  \left[\Phi(z_c)-z_c\phi(z_c)+\frac{\phi^2(z_c)}{1-\Phi(z_c)}\right].
\end{align*}
Therefore, with no censoring in training data, the proposed ICMSE criterion is:
\begin{align}
   \text{ICMSE}(\bm{x}_{n+1})=\int_{[0,1]^p}\sigma_{\rm new}^2-\sigma_{\rm new}^2\rho^2  \left[\Phi(z_c)-z_c\phi(z_c)+\frac{\phi^2(z_c)}{1-\Phi(z_c)}\right]\; d\bm{x}_{\rm new}.
   \label{AppEqu:CMSEwithcensor}
\end{align}

\subsection{Proof of Proposition 2}\label{Sec:AppCMSEc}
Consider a more general case \textit{with} censoring in training data, i.e., $\mathcal{Y}_n = \{\bm{y}_o,\bm{y}_c' \geq\bm{c}\}$. It is important to note that, due to the existence of censoring data $\{\bm{y}_c' \geq\bm{c}\}$, the random variable $\xi(\bm{x}_{\rm new})|\mathcal{Y}_n$ is \textit{no longer} normally distributed. This, in turn, requires more cumbersome derivations than the earlier case without censoring in training data.

Using the conditional mean and variance expressions (2.3) and (2.4) for the \textit{censored} GP, we have 
\begin{align}
\begin{split}
\mathbb{E}\left[\xi(\bm{x}_{\rm new})|Y_{n+1}'\geq c\right]& =\mu_\xi + \boldsymbol{\gamma}_{n+1,\rm new}^T \bm{\Gamma}_{n+1}^{-1}\left([\bm{y}_{o}, \hat{\bm{y}}_{c},\hat{y}_{n+1}^{(> )}]^T - \mu_\xi \cdot \textbf{1}_{n+1}\right), \\ 
\mathbb{E}\left[\xi(\bm{x}_{\rm new})|Y_{n+1}'< c\right] & =\mu_\xi + \boldsymbol{\gamma}_{n+1,\rm new}^T \bm{\Gamma}_{n+1}^{-1}\left([\bm{y}_{o}, \hat{\bm{y}}_{c},\hat{y}_{n+1}^{(<)}]^T - \mu_\xi \cdot \textbf{1}_{n+1}\right), \\
    \text{Var}[\xi(\bm{x}_{\rm new})|Y_{n+1}'\geq c]   &  = \sigma^2_\xi-\boldsymbol{\gamma}_{n+1,\rm new}^T \bm{\Gamma}_{n+1}^{-1}\left(\bm{\Gamma}_{n+1}-\bm{\Sigma}_1\right)\bm{\Gamma}_{n+1}^{-1}\boldsymbol{\gamma}_{n+1,\rm new}, \\
       \mathbb{E}_{Y_{n+1}|Y_{n+1}'<c}[\text{Var}(\xi(\bm{x}_{\rm new})|Y_{n+1})]  &  = \sigma^2_\xi-\boldsymbol{\gamma}_{n+1,\rm new}^T \bm{\Gamma}_{n+1}^{-1}\left(\bm{\Gamma}_{n+1}-\hat{\bm{\Sigma}}\right)\bm{\Gamma}_{n+1}^{-1}\boldsymbol{\gamma}_{n+1,\rm new},
    \label{AppEqu:CMSETrems}
\end{split}
\end{align}
where $\mu_\xi$ and $\sigma^2_\xi$ are the mean and variance for the prior GP, respectively. Here, $\boldsymbol{\gamma}_{n+1,\rm new}=\sigma^2_\xi\big[R_{\boldsymbol{\theta}_\xi}(\bm{x}_1, \bm{x}_{\rm new}), \cdots, R_{\boldsymbol{\theta}_\xi}(\bm{x}_{n+1}, \bm{x}_{\rm new}) \big]^T$ and $\bm{\Gamma}_{n+1} = \sigma^2_\xi{[R_{\boldsymbol{\theta}_\xi}(\bm{x}_i, \bm{x}_j)]_{i=1}^{n+1}} _{j=1}^{n+1} + \sigma^2_\epsilon \bm{I}_{n+1}$. Furthermore, $\hat{y}_{n+1}^{(> )} = \mathbb{E}(Y_{n+1}'|\bm{y}_o,Y_{n+1}'\geq c)$ and  $\hat{y}_{n+1}^{(<)} = \mathbb{E}(Y_{n+1}'|\bm{y}_o,Y_{n+1}'<c)$ are the expected response for the potential observation, $\bm{\Sigma}_1=\bm{\Sigma}_1(\bm{x}_{n+1})=\text{diag}\left(\bm{0}_{n_o},\text{Var}([\bm{y}_c',Y_{n+1}]|\bm{y}_{o},\bm{y}_c' \geq\bm{c},Y_{n+1}=c)\right)$, and $\hat{\bm{\Sigma}}=\hat{\bm{\Sigma}}(\bm{x}_{n+1})=\text{diag}\left(\bm{0}_{n_o},\mathbb{E}_{Y_{n+1}|\bm{y}_{o}}[\text{Var}(\bm{y}_c'|Y_{n+1},\bm{y}_{o},\bm{y}_c' \geq\bm{c})],0 \right)$.  Plugging in equation \eqref{AppEqu:CMSETrems} back into \eqref{AppEqu:CMSE}, we have 
\begin{align*}
\begin{split}
     \text{CMSE}
     &=\sigma^2_{\rm new}-\lambda(1-\lambda)\left( \boldsymbol{\gamma}_{n+1,\rm new}^T \bm{\Gamma}_{n+1}^{-1}[\bm{0}_n,\hat{y}_{n+1}^{(> )}-\hat{y}_{n+1}^{(<)}]^T \right)^2 \\
     & - (1-\lambda) \boldsymbol{\gamma}_{n+1,\rm new}^T \bm{\Gamma}_{n+1}^{-1}\left(\bm{\Sigma}_{1}-\hat{\bm{\Sigma}}\right)\bm{\Gamma}_{n+1}^{-1}\boldsymbol{\gamma}_{n+1,\rm new}.     
\end{split}
\end{align*}
Here, $\sigma^2_{\rm new}=\sigma^2_\xi-\boldsymbol{\gamma}_{n,\rm new}^T \bm{\Gamma}^{-1}_n\boldsymbol{\gamma}_{n,\rm new} +\boldsymbol{\gamma}_{n,\rm new}^T \bm{\Gamma}_n^{-1}\hat{\bm{\Sigma}}\bm{\Gamma}_n^{-1}\boldsymbol{\gamma}_{n,\rm new}$ and the probability of censoring
$\lambda=\mathbb{P}(\bm{y}_{c}\geq \bm{c}, Y_{n+1}'\geq c|\bm{y}_{o})/\mathbb{P}(\bm{y}_{c}\geq \bm{c}|\bm{y}_{o})$. (The computation of these orthant probabilities and moments of
the truncated multivariate normal distribution will be discussed later in Appendix \ref{Sec:appImp}.) Putting everything together, our ICMSE criterion has the following explicit form:
\begin{align}
     &\text{ICMSE}
     =\int_{[0,1]^p}\sigma^2_{\rm new}-\boldsymbol{\gamma}_{n+1,\rm new}^T \bm{\Gamma}_{n+1}^{-1}\bm{H}_c(\bm{x}_{n+1})\bm{\Gamma}_{n+1}^{-1}\boldsymbol{\gamma}_{n+1,\rm new}\;d\bm{x}_{\rm new}
    \label{AppEqu:SFH}
\end{align}
where $\bm{H}_c(\bm{x}_{n+1}) =  (1-\lambda)\left(\bm{\Sigma}_{1}-\hat{\bm{\Sigma}}\right)+\lambda (1-\lambda)\text{diag}\left(\bm{0}_n,\hat{y}_{n+1}^{(> )}\hat{y}_{n+1}^{(<)}\right)$.

\section{Bi-fidelity ICMSE design criterion}
\subsection{Proof of Proposition 3} \label{Sec:appMFICMES}
For the bi-fidelity setting, the training data is $\{\bm{f},\mathcal{Y}_n\}$. Using the conditional mean and variance expressions (3.3) and (3.4) for the \textit{bi-fidelity} GP model, we have   
\begin{align}
\begin{split}
\mathbb{E}\left[\xi(\bm{x}_{\rm new})|Y_{n+1}'\geq c\right]& =\mu_f + \boldsymbol{\gamma}_{n+1,\rm new}^T \bm{\Gamma}_{n+1}^{-1}\left([\bm{f},\bm{y}_{o}, \hat{\bm{y}}_{c},\hat{y}_{n+1}^{(> )}]^T - \mu_f \cdot \textbf{1}_{n+1}\right), \\ 
\mathbb{E}\left[\xi(\bm{x}_{\rm new})|Y_{n+1}'< c\right] & =\mu_f + \boldsymbol{\gamma}_{n+1,\rm new}^T \bm{\Gamma}_{n+1}^{-1}\left([\bm{f},\bm{y}_{o}, \hat{\bm{y}}_{c},\hat{y}_{n+1}^{(<)}]^T - \mu_f \cdot \textbf{1}_{n+1}\right), \\
    \text{Var}[\xi(\bm{x}_{\rm new})|Y_{n+1}'\geq c]   &  = \sigma^2_f+\sigma^2_\delta-\boldsymbol{\gamma}_{n+1,\rm new}^T \bm{\Gamma}_{n+1}^{-1}\left(\bm{\Gamma}_{n+1}-\bm{\Sigma}_1\right)\bm{\Gamma}_{n+1}^{-1}\boldsymbol{\gamma}_{n+1,\rm new},\\
       \mathbb{E}_{Y_{n+1}|Y_{n+1}'<c}[\text{Var}(\xi(\bm{x}_{\rm new})|Y_{n+1})]  &  = \sigma^2_f+\sigma^2_\delta-\boldsymbol{\gamma}_{n+1,\rm new}^T \bm{\Gamma}_{n+1}^{-1}\left(\bm{\Gamma}_{n+1}-\hat{\bm{\Sigma}}\right)\bm{\Gamma}_{n+1}^{-1}\boldsymbol{\gamma}_{n+1,\rm new}.
    \label{AppEqu:Fusion_CMSETrems} 
\end{split}
\end{align}
Though equations \eqref{AppEqu:Fusion_CMSETrems} appears to be quite similar to the equations \eqref{AppEqu:CMSETrems}, the notations in \eqref{AppEqu:Fusion_CMSETrems} are overloaded with the bi-fidelity expressions (3.3) and (3.4) for simplicity (see Section 3.1). Here, $\mu_f$ and $\sigma^2_f$ are the mean and variance of the GP $f(\cdot)$ modeling the computer experiment, 
$\boldsymbol{\gamma}_{n+1,\rm new}=\sigma^2_f[R_{\boldsymbol{\theta}_f}(\bm{x}_i, \bm{x}_{\rm new})]_{i=1}^{n+1}+\sigma^2_\delta[\bm{0}_{n-m},R_{\boldsymbol{\theta}_\delta}(\bm{x}_i, \bm{x}_{\rm new})]_{i=1}^{m+1}$,
and $\bm{\Gamma}_{n+1}=\sigma^2_f{[R_{\boldsymbol{\theta}_f}(\bm{x}_i, \bm{x}_j)]_{i=1}^{n+1}}_{j=1}^{n+1} + \text{diag}\left( \bm{0}_{n-m},\sigma^2_\delta{[R_{\boldsymbol{\theta}_\delta}(\bm{x}_i, \bm{x}_j)]_{i=1}^{m+1}}_{j=1}^{m+1} + \sigma^2_\epsilon \bm{I}_{m+1} \right)$. Furthermore, $\hat{y}_{n+1}^{(> )} = \mathbb{E}(Y_{n+1}'|\bm{f},\bm{y}_o,Y_{n+1}'\geq c)$ and  $\hat{y}_{n+1}^{(<)} = \mathbb{E}(Y_{n+1}'|\bm{f},\bm{y}_o,Y_{n+1}'<c)$ are the expected responses for the potential observation, $\bm{\Sigma}_1=\bm{\Sigma}_1(\bm{x}_{n+1})=\text{diag}\left(\bm{0}_{n-n_c},\text{Var}(\bm{y}_c' \geq\bm{c},Y_{n+1}=c|\bm{f},\bm{y}_{o})\right)$, and $\hat{\bm{\Sigma}}=\hat{\bm{\Sigma}}(\bm{x}_{n+1})=\text{diag}\left(\bm{0}_{n-n_c},\mathbb{E}_{Y_{n+1}|\bm{f},\bm{y}_{o}}[\text{Var}(\bm{y}\geq \bm{c}|Y_{n+1},\bm{f},\bm{y}_{o},\bm{y}_c' \geq\bm{c})],0 \right)$.

Plugging in \eqref{AppEqu:Fusion_CMSETrems} to \eqref{AppEqu:CMSE}, we have the following explicit form ICMSE criterion:
\begin{align}
     \text{ICMSE} (\bm{x}_{n+1})
     &=\int_{[0,1]^p}\sigma^2_{\rm new}-\boldsymbol{\gamma}_{n+1,\rm new}^T \bm{\Gamma}_{n+1}^{-1}\bm{H}_c(\bm{x}_{n+1})\bm{\Gamma}_{n+1}^{-1}\boldsymbol{\gamma}_{n+1,\rm new}\;d\bm{x}_{\rm new},    
\end{align}
where, $\sigma^2_{\rm new}=\sigma^2_f+\sigma^2_\delta-\boldsymbol{\gamma}_{n,\rm new}^T \bm{\Gamma}^{-1}_n\boldsymbol{\gamma}_{n,\rm new} +\boldsymbol{\gamma}_{n,\rm new}^T \bm{\Gamma}_n^{-1}\hat{\bm{\Sigma}}\bm{\Gamma}_n^{-1}\boldsymbol{\gamma}_{n,\rm new}$, and
\begin{align}
    \bm{H}_c(\bm{x}_{n+1}) =  (1-\lambda)\left(\bm{\Sigma}_{1}-\hat{\bm{\Sigma}}\right)+\lambda (1-\lambda)\text{diag}\left(\bm{0}_n,\hat{y}_{n+1}^{(> )}\hat{y}_{n+1}^{(<)}\right).
    \label{AppEqu:MFH}
\end{align}
Here, the probability of censoring
$\lambda=\mathbb{P}(\bm{y}_{c}\geq \bm{c}, Y_{n+1}'\geq c|\bm{f},\bm{y}_{o})/\mathbb{P}(\bm{y}_{c}\geq \bm{c}|\bm{f},\bm{y}_{o})$.

\subsection{Proof of Corollary 1} \label{Sec:appSimpPC}
Note that in ICMSE criterion (3.6), $\boldsymbol{\gamma}_{n+1,\rm new}$ is a function of both $\bm{x}_{n+1}$ and $\bm{x}_{\rm new}$, $\bm{\Gamma}_{n+1}$ and $\bm{H}_c(\bm{x}_{n+1})$ are only function of $\bm{x}_{n+1}$, and $\sigma^2_{\rm new}$ is only a function of $\bm{x}_{\rm new}$;  it can therefore be further simplified as
\begin{align*}
     \text{ICMSE} (\bm{x}_{n+1}) = \bar{\sigma}^2- \text{tr} \left(\bm{\Gamma}_{n+1}^{-1}\bm{H}_c(\bm{x}_{n+1})\bm{\Gamma}_{n+1}^{-1}
    \bm{\Lambda}\right),    
\end{align*}
where $\bar{\sigma}^2 = \int \sigma^2_{\rm new} \; d\bm{x}_{\rm new}$ is a constant with respect to
$\bm{x}_{n+1}$, $\text{tr(A)}=\sum_i A_{i,i}$ is the trace of matrix A, and $\bm{\Lambda} = \int \boldsymbol{\gamma}_{n+1,\rm new}^T\boldsymbol{\gamma}_{n+1,\rm new}\;d\bm{x}_{\rm new}$. Assume the following product correlation structure:
\begin{align}
R_{\boldsymbol{\theta}_f} (\bm{x},\bm{x}')=\prod_{l=1}^p R_{\boldsymbol{\theta}_f}^{(l)} (x_l,x_l'), \quad R_{\boldsymbol{\theta}_\delta} (\bm{x},\bm{x}')=\prod_{l=1}^p R_{\boldsymbol{\theta}_\delta}^{(l)} (x_l,x_l'),
\end{align}
and denote $ \zeta^{(l)}(z,x) = R_{\boldsymbol{\theta}_f}^{(l)} (z,x) +\mathds{1}_{\{i\geq (n-m)\}}R_{\boldsymbol{\theta}_\delta}^{(l)} (z,x)$. We can simplify the $p$-dimensional integral in $\bm{\Lambda}$ to a product of $1$-dimensional integrals:
\begin{align}
\Lambda_{ij}= \int_{[0,1]^p} \prod_{l=1}^p 
\zeta^{(l)}(x_{i,l},x_l) \zeta^{(l)}(x_{j,l},x_l)\; d\bm{x} = \prod_{l=1}^p \left[\int_0^1 
\zeta^{(l)}(x_{i,l},x_l) \zeta^{(l)}(x_{j,l},x_l)\; dx_l \right],
\label{AppEqu:LambdaInt}
\end{align}
where $i,j =1,2,\cdots, n+1$.

Furthermore, under the following product \textit{Gaussian} correlation structure:
\begin{align}
R_{\boldsymbol{\theta}_f} (\bm{x},\bm{x}')=\prod_{l=1}^p \theta_{f,\;l} ^{4(x_{i}-x_{i}')^2}, \quad R_{\boldsymbol{\theta}_\delta} (\bm{x},\bm{x}')=\prod_{l=1}^p\theta_{\delta, \;l} ^{4(x_{i}-x_{i}')^2},
\label{AppEqu:GaussanCorr}
\end{align}
the $1$-dimensional integrals of entries in $\bm{\Lambda}$ \eqref{AppEqu:LambdaInt} can be reduced to integrals of exponential polynomial expressions, which have an easy-to-evaluate form. 
Consider the following general expression for an exponential polynomial: 
\begin{align*}
G([a,x], [b,y]) &=\int_{0}^1 \exp \left[ - a (x-z)^2\right]\exp \left[ - b (y-z)^2\right] dz\\
&=\sqrt{\frac{\pi}{a+b}}\exp \left(\frac{(ax+by)^2}{a+b}-ax^2-by^2\right) \\
&\times \left[\Phi\left( \sqrt{\frac{2}{a+b}} \left( a+b-ax-by\right)\right) -\Phi\left( -\sqrt{\frac{2}{a+b}} \left( ax+by\right)\right)\right],
\end{align*}
where $\Phi(\cdot)$ is the CDF for standard normal. Under \eqref{AppEqu:GaussanCorr}, the entries of $\bm{\Lambda}$ \eqref{AppEqu:LambdaInt} can be further simplified as:
\begin{align}
\begin{split}
\Lambda_{ij}=&
 \prod_{l=1}^p G([\tilde{\theta}_{f,l},{x}_{i,l}],[ \tilde{\theta}_{f,l},{x}_{j,l}]) +\mathds{1}_{\{i\geq (n-m)\}}\mathds{1}_{\{j\geq (n-m)\}} \prod_{l=1}^p G([\tilde{\theta}_{\delta,l},{x}_{i,l}],[ \tilde{\theta}_{\delta,l},{x}_{j,l}])  \notag\\
+&\mathds{1}_{\{i\geq (n-m)\}} \prod_{l=1}^p G([\tilde{\theta}_{f,l},{x}_{i,l}],[\tilde{\theta}_{\delta,l},{x}_{j,l}])+\mathds{1}_{\{j\geq (n-m)\}} \prod_{l=1}^p G([\tilde{\theta}_{\delta,l},{x}_{i,l}],[\tilde{\theta}_{f,l},{x}_{j,l}]),
\end{split}\label{AppEqu:Int}
\end{align}
where $\tilde{\boldsymbol{\theta}_{f}}=-4\log\boldsymbol{\theta}_{f}$ and $\tilde{\boldsymbol{\theta}_{\delta}}=-4\log\boldsymbol{\theta}_{\delta}$.

Note that the above simplification under product Gaussian correlations can also be used in the single-fidelity ICMSE criterion (2.11) as well, in which case $\Lambda_{ij}=\prod_{l=1}^p G([\tilde{\theta}_{\xi,l},{x}_{i,l}],[ \tilde{\theta}_{\xi,l},{x}_{j,l}])$, with $\tilde{\boldsymbol{\theta}_{\xi}}=-4\log\boldsymbol{\theta}_{\xi}$.

\section{Computational approximations} \label{Sec:appImp}
In practice, the computation of the matrix $\hat{\boldsymbol{\Sigma}}$ in the single-fidelity expression \eqref{AppEqu:CMSETrems} or the bi-fidelity expression \eqref{AppEqu:Fusion_CMSETrems} can be quite time-consuming, since a closed-form expression is difficult to obtain for the expected variance term (the conditional distributions $[Y_{n+1}|\bm{y}_o]$ and $[Y_{n+1}|\bm{f},\bm{y}_o]$ are non-Gaussian). For the single-fidelity setting, we found the following approximation to be useful for efficient computation:
\begin{align*}
\mathbb{E}_{Y_{n+1}<c|\mathcal{Y}_n}[\text{\rm Var}(\bm{y}_{c}'|\bm{y}_{o},\bm{y}_{c}'\geq \bm{c},Y_{n+1})]\approx \text{Var}(\bm{y}_{c}'|\bm{y}_{o},\bm{y}_{c}'\geq \bm{c},Y_{n+1} = \hat{Y}_{n+1}),    
\end{align*}
where $\hat{Y}_{n+1}=\mathbb{E}(Y_{n+1}|\mathcal{Y}_n,Y_{n+1}<c)$. Similar simplification also applies for the bi-fidelity setting.  This can be viewed as a plug-in estimate (with $Y_{n+1} = \hat{Y}_{n+1}$) which approximates the conditional mean expression on the left-hand side. The right-hand approximation can be efficiently computed via the truncated moments of a multivariate normal distribution, as implemented in the R package \texttt{tmvtnorm}.

\end{document}